\documentclass[english,prb,twocolumn,aps]{revtex4-1}
\usepackage[T1]{fontenc}
\usepackage[latin9]{inputenc}
\usepackage{amsmath}
\usepackage{amssymb}
\usepackage{graphicx}
\usepackage{esint}

\makeatletter

\@ifundefined{textcolor}{}
{%
 \definecolor{BLACK}{gray}{0}
 \definecolor{WHITE}{gray}{1}
 \definecolor{RED}{rgb}{1,0,0}
 \definecolor{GREEN}{rgb}{0,1,0}
 \definecolor{BLUE}{rgb}{0,0,1}
 \definecolor{CYAN}{cmyk}{1,0,0,0}
 \definecolor{MAGENTA}{cmyk}{0,1,0,0}
 \definecolor{YELLOW}{cmyk}{0,0,1,0}
}

\makeatother

\usepackage{babel}
\begin{document}

\title{Open system quantum annealing in mean field models with exponential
degeneracy}

\author{Kostyantyn Kechedzhi$^{1}$ and Vadim N. Smelyanskiy$^{2}$}

\affiliation{$^{1}$QuAIL, NASA Ames Research Center, Mail Stop 269-3, Moffett
Field, CA 94035 and \\
USRA, NASA Ames Research Center, Moffett Field, CA 94035 \\
 $^{2}$Google, 150 Main St, Venice Beach, CA, 90291}
\begin{abstract}
Real life quantum computers are inevitably affected by intrinsic noise
resulting in dissipative non-unitary dynamics realized by these devices.
We consider an open system quantum annealing algorithm optimized for
a realistic analog quantum device which takes advantage of noise-induced
thermalization and relies on incoherent quantum tunneling at finite
temperature. We analyze the performance of this algorithm considering
a $p-$spin model which allows for a mean field quasicalssical solution
and at the same time demonstrates the 1st order phase transition and
exponential degeneracy of states. We demonstrate that finite temperature
effects introduced by the noise are particularly important for the
dynamics in presence of the exponential degeneracy of metastable states.
We determine the optimal regime of the open system quantum annealing
algorithm for this model and find that it can outperform simulated
annealing in a range of parameters. 
\end{abstract}
\maketitle

\section{Introduction}

Quantum computing hardware is affected by substantial level of intrinsic
noise and therefore naturally realizes dissipative quantum dynamics~\cite{Leggett,NielsenChuang2011}.
Optimization algorithms, where a configuration of a binary string
$x$ minimizing a given (energy or cost) function $f(x)$ is sought
for, naturally extract computational advantage from the irreversible
dissipative dynamics, and could therefore be readily implemented on
a number of existing hardware platforms~\cite{SmelyanskiyGoogle14,britton_engineered_2012}.
More specifically, quantum annealing~\cite{PhysRevE.58.5355} (QA)
is a quantum analog of the widely applied classical simulated annealing
algorithm~\cite{GemanGeman} (SA), a heuristic solver of NP-hard
optimization problems~\cite{Finnila1994,PhysRevE.58.5355,DisorderedMagnetScience99,Farhi2002,SantoroCarr2002},
with quantum fluctuations playing the role analogous to thermal fluctuations
in simulated annealing. NP-hard optimization problems such as finding
a ground state spin configuration of a spin glass, are often characterized
by an energy landscape with a large number of local minima separated
by extensive energy barriers. Dissipative dynamics realized by the
open system quantum annealing provides an efficient mechanism for
thermalization within domains of attraction of local minima. For efficient
search of the configuration space the barriers separating different
domains of attraction have to be overcome which may proceed via thermal
excitation or quantum tunneling process. The performance of the open
system quantum annealing algorithm is therefore characterized by a
set of relaxation rates associated with such processes, as opposed
to, for example, the spectral gaps, as is the case for adiabatic quantum
algorithm~\cite{Farhi2002,Bapst2013127}. 

The longest relaxation times correspond to the often exponentially
slow transitions between local minima separated by extensive potential
barriers. Unitary dynamics of a pair of such states corresponds to
the switching rate of the order of the matrix element $\Delta$, which
in presence of an extensive barrier may scale exponentially with the
system size $\Delta\propto\exp(-\mathrm{const}\times N)$. Whereas
fast dissipative relaxation within a domain of attraction of a local
minima due to the hardware noise introduces a lifetime or level width
$W$. This fast local relaxation strongly suppresses the coherent
superposition of the states localized in different local minima when
$W\gg\Delta$. Nevertheless, the incoherent quantum tunneling is possible
in presence of such strong dissipation where the transition rate is
described by the Fermi golden rule type expression $\propto\Delta^{2}\ll\Delta$.
This is the regime likely realized in a large scale quantum annealer~\cite{SmelyanskiyGoogle14}.
It is an open question whether such incoherent extensive quantum tunneling
may provide a more efficient mechanism for searching the configuration
space as compared to classical simulated annealing relying on thermal
excitation.

\emph{}

In this paper\emph{,} we analyze the performance of the open system
quantum annealing algorithm optimized for the regime of the quantum
dissipative dynamics, taking advantage of thermalization and incoherent
quantum tunneling. Our goal here is to analyze the contribution of
extensive quantum tunneling to the performance of the algorithm. We
consider a model in which there exists a metastable state separated
by an extensive barrier from the ground state. We consider a system
of Ising spins interacting each with each other with $p-$body interaction
of equal strength, a model often referred to as a $p-$spin model.
This model allows for a quasiclassical WKB description~\cite{BraunRMP,Owerre20151},
where the expansion is performed in $1/N$ rather than the more usual
$\hbar$, and, at the same time, demonstrates key features characteristic
of a range of complex (NP-hard) optimization problems, such as the
1st order phase transition (for $p\geq3$) and exponentially small
gap between the ground and excited state. Crucially, the metastable
state realized in this model is characterized by an exponential degeneracy
whereas the ground state is unique. Such entropic imbalance is in
fact typical for low energy states in the spin glass phase and it
strongly affects the low temperature system dynamics in both quantum
and classical cases. The effect of entropic imbalance is the main
focus of our analysis in this paper. 

We demonstrate that in presence of such extensive entropy imbalance
SA computation time scales exponentially with the system size $N$.
This can be understood intuitively considering the performance of
the simulated annealing applied to a model demonstrating a 1st order
phase transition into a state characterized by an order parameter.
Assume the ordered state to be a ferromagnet for simplicity. In simulated
annealing the system is initialized at infinite temperature, or equal
occupation of all classical spin states, then the temperature is gradually
lowered to zero. The simulated spin dynamics is chosen to satisfy
the detailed balance condition such that it samples the thermal distribution
at a given (instant) temperature. The initial state is a paramagnet,
and therefore the solution, the ground state spin configuration at
zero temperature, is expected to have high statistical weight only
at low enough temperatures below the ferromagnetic phase transition.
The exponential degeneracy of the metastable state corresponds to
the entropy linear in the system size $N$ which significantly lowers
the transition temperature. This can be understood intuitively from
the following argument. We assume a mean field case in which energies
of the metastable and ground states as well as the barrier separating
them scale linearly with the system size $NE_{MS}$, $NE_{GS}$ and
$NU$, respectively. We find from equating the free energies $NQ_{MS}-NE_{MS}/T\approx-NE_{GS}/T$,
where the entropy imbalance is given by $NQ_{MS}$, that the transition
occurs at $T_{c}\sim\frac{E_{MS}-E_{GS}}{Q_{MS}}\sim O(1)$. Furthermore,
high statistical weight of the ground state is achieved only after
equilibration at the low temperature below the phase transition $T\lesssim T_{c}\sim O(1)$.
In presence of the extensive barrier $NU$ the relaxation towards
the thermal distribution described by the classical Krammers escape
rate $\sim\exp\left(-NU/T_{c}\right)$ is exponentially slow. Moreover,
the entropy gradient along the over-the-barrier escape trajectory
gives rise to an additional entropic factor $\sim\exp(Q_{MS}-Q_{T})$,
where $Q_{T}$ is the entropy corresponding to the top of the barrier,
which appears as an exponential contribution to the prefactor of the
Krammers rate~\cite{HanggiRMP}. Therefore the computation time allowing
for such relaxation to occur is at least as long the relaxation time
$\tau_{u}\sim\exp(NUQ_{MS}/(E_{MS}-E_{GS})+Q_{MS}-Q_{T})$. In fact,
a more careful analysis, see Appendix~\ref{sec:Simulated-Annealing},
shows that the optimal SA computation time in the presence of the
extensive entropy imbalance is given by the smallest of either $\tau_{u}$
or the exhaustive search time $\tau_{es}\sim2^{N}$. At the same time
quantum tunneling amplitude saturates as $T\rightarrow0$ and may
be more efficient than over the barrier escape suggesting that quantum
annealing could be more efficient than simulated annealing. Note however
that quantum tunneling rate in this mean field model will also scale
exponentially with $N$. Therefore the performance of each algorithm
will be characterized by a numerical factor in the exponent which
have to be carefully compared. The result is not obvious a priori
since we are comparing here different microscopic mechanisms: the
quantum dynamics constrained by conservation laws with the classical
thermal excitation process constrained by the entropy imbalance and
the low temperature.

Considering the open system quantum annealing applied to the $p$-spin
model we show that the scaling of the optimal QA computation time
(allowing for repeated runs of the algorithm) is determined by the
quantum tunneling amplitude at a single point in the algorithm, the
so called freezing point, after which quantum/thermal fluctuations
are weak and the transitions over or through the barrier are no longer
likely. We find that due to the effect of the entropy associated with
the metastable state the optimal quantum tunneling rate is achieved
at vanishing temperature, i.e. raising temperature may reduce the
quantum tunneling rate. This is in contrast with the usual intuition
about a quantum mechanical particle trapped in a non-degenerate metastable
potential well where the escape rate monotonously increases with temperature.
The optimal QA regime therefore also corresponds to the vanishing
temperature. Comparing the optimal computation time of QA obtained
in this regime with that of SA for a range of the potential barrier
shapes we find that QA could outperform SA under certain circumstances,
thus providing a polynomial (rather than exponential) speedup.

The remainder of the paper is organized as follows. In Sec.~\ref{sec:The-model}
we introduce the $p$-spin model and its WKB analysis describing the
evolution of the potential energy and the transition rates in the
course of the quantum annealing algorithm. In Sec.~\ref{sec:Computation-time}
we discuss the dynamics of the model in the course of the quantum
annealing and identify the freezing point and its optimal position
in the course of the algorithm. We conclude with a discussion of the
results in Sec.~\ref{sec:Discussion}.

\section{The model\label{sec:The-model}}

We consider $N$ Ising spins $1/2$ on a fully connected graph (i.e.
each spin interacts with each other spin) with uniform interaction
strength such that the system is fully described in terms of the total
spin projection operator $\hat{S}^{\alpha}=\sum_{i=1}^{N}\hat{\sigma}_{i}^{\alpha}$,
where $\alpha=x,y,z$ and $\hat{\sigma}^{\alpha}$ is a set of spin$-1/2$
operators. We consider a Hamiltonian,

\begin{equation}
\mathcal{H}=sNf\left(\frac{2}{N}\hat{S}^{z}\right)-(1-s)\hat{S}^{x},\label{eq:Hamiltonian}
\end{equation}
consisting of a uniform transverse field term, the second term above,
and a potential energy of interaction $f(x)$, which is assumed to
be a function of the $z-$projection operator. The potential energy
of a $p$-spin interaction of unit strength $H=\left(\frac{2}{N}\right)^{p}\sum\hat{\sigma}_{i_{1}}^{z}\hat{\sigma}_{i_{2}}^{z}...\hat{\sigma}_{i_{p}}^{z}$
corresponds to $f(x)=x^{p}$. Without loss of generality we choose
both of the terms in the Hamiltonian Eq.~(\ref{eq:Hamiltonian})
to scale linearly with $N$. The parameter $s$ in Eq.~(\ref{eq:Hamiltonian})
controls the relative strength of the potential energy and the transverse
field and changes from $s=0$ to $s=1$ in the course of the quantum
annealing algorithm. 

Of specific interest is the case of $p=3$. Models with $p>3$ can
be solved in polynomial time by avoiding the 1st order phase transition
using advanced driver Hamiltonian, $b-$spin ($b\geq2$) interactions
inducing transverse (XY) ferro- or antiferromagnetism, in addition
to the standard transverse field term~\cite{PhysRevE.85.051112,Nishimori2012,PhysRevE.75.051112}.
In this case arriving at the ground state at the end of the QA evolution
does not require extensive quantum tunneling through a barrier. Therefore
a classical model algorithm such as ``spin-vector'' Monte Carlo
method~\cite{SSSV}, which could follow the deformation of the effective
quasiclassical potential imitating the course of QA, is also not expected
to encounter a bottleneck requiring exponential computation time.
Whereas in the case of $p=3$ the 1st order phase transition cannot
be avoided in this way, as a result an exponentially weak tunneling
process is critical for finding the ground state in this system. The
quantum tunneling regime in this case cannot be described by a classical
model. While adiabatic quantum computation was analyzed in the p-spin
model~\cite{Jorg2010,SemerjianPspin}, we will focus here on the
open system quantum annealing in presence of dissipation and non-zero
temperature which is the case more suitable for implementation on
current analog quantum annealers. 

The Hamiltonian Eq.~(\ref{eq:Hamiltonian}) commutes with $\hat{S}^{2}\equiv(\hat{S}^{x})^{2}+(\hat{S}^{y})^{2}+(\hat{S}^{z})^{2}$,
which is therefore a conserved quantity. In the basis of states, $|S,M\rangle:$~$\hat{S}^{2}|S,M\rangle=S(S+1)|S,M\rangle,$
with definite total spin $S$ and its projection on $z$-axis $M=\left\{ -S,...,S\right\} $
the matrix elements of the Hamiltonian Eq.~(\ref{eq:Hamiltonian})
are given by the standard spin$-S$ rules,

\begin{eqnarray}
 & \hat{S}^{z}|S,M\rangle=M|S,M\rangle\\
 & \hat{S}^{\pm}|S,M\rangle=\sqrt{S(S+1)-M(M\pm1)}|S,M\pm1\rangle\label{eq:MatrixElemnt}
\end{eqnarray}
where we introduced raising and lowering operators, $\hat{S}^{\pm}=\frac{1}{2}(\hat{S}^{x}\pm i\hat{S}^{y})$.
We introduce an integer parameter $K=0,1,...,\left\lfloor \frac{N}{2}\right\rfloor $
to label the total spin eigenstates $S=\frac{N}{2}-K$, each value
corresponding to a completely disconnected subspace of the eigenspace
of Eq.~(\ref{eq:Hamiltonian}), which will however be connected due
to the coupling to a thermal bath. The Hamiltonian Eq.~(\ref{eq:Hamiltonian})
is symmetric with respect to exchanges of pairs of spins $\hat{\sigma}_{i}\leftrightarrow\hat{\sigma}_{j}$,
and in fact with respect to all permutations of spins, since such
operations do not change the sum over all spins $\hat{S}^{\alpha}$.
This symmetry introduces a high degeneracy of eigenstates depending
on their total spin $S$. The subspace with the maximal total spin
$S=\frac{N}{2}$ or $K=0$ contains $2S+1$ non-degenerate states
(there are no non-trivial permutations) corresponding to all possible
projections of the total spin on $z-$axis. At the same time the states
with $K\neq0$ are highly degenerate, with the degeneracy being determined
by the representations of the group of permutations. The eigenstate
with a total spin labeled by $K$ has the degeneracy $(\begin{smallmatrix}N\\
K
\end{smallmatrix})-(\begin{smallmatrix}N\\
K-1
\end{smallmatrix})\sim\exp\left(NQ_{k}\right)$, where $k\equiv\frac{K}{N}=\left\{ 0,\frac{1}{N},\frac{2}{N},...,\frac{1}{N}\left\lfloor \frac{N}{2}\right\rfloor \right\} $
, which corresponds to the entropy term, 
\[
Q_{k}\approx-k\ln k-(1-k)\ln(1-k)+O(\frac{\ln N}{N}),
\]
that has to be added to the free energy of a state with a given energy
$E$ and total spin parameter $k$, $\mathcal{F}=\beta E+Q_{k}$. 

In this paper we are interested in a cubic potential energy,

\begin{equation}
f(q)=-c\left(q-q_{min}\right)^{2}\left(q-\frac{3q_{max}-q_{min}}{2}\right),\label{eq:CubicPotential}
\end{equation}
where $q\equiv\frac{2}{N}M$, $q=\left\{ -(1-2k),...,(1-2k)\right\} $.
Eq.~(\ref{eq:CubicPotential}) is the most general cubic function
with the metastable minimum at $q_{min}$ and the potential barrier
top at $q_{max}$, where $f(q_{min})>f(1)$ ensures that $q=1$ is
the global minimum. Without loss of generality we can put $c=1$,
the only effect of $c\neq1$ is to rescale the parameter $s$ in Eq.~(\ref{eq:Hamiltonian}).

\begin{figure}
\includegraphics[width=0.32\columnwidth]{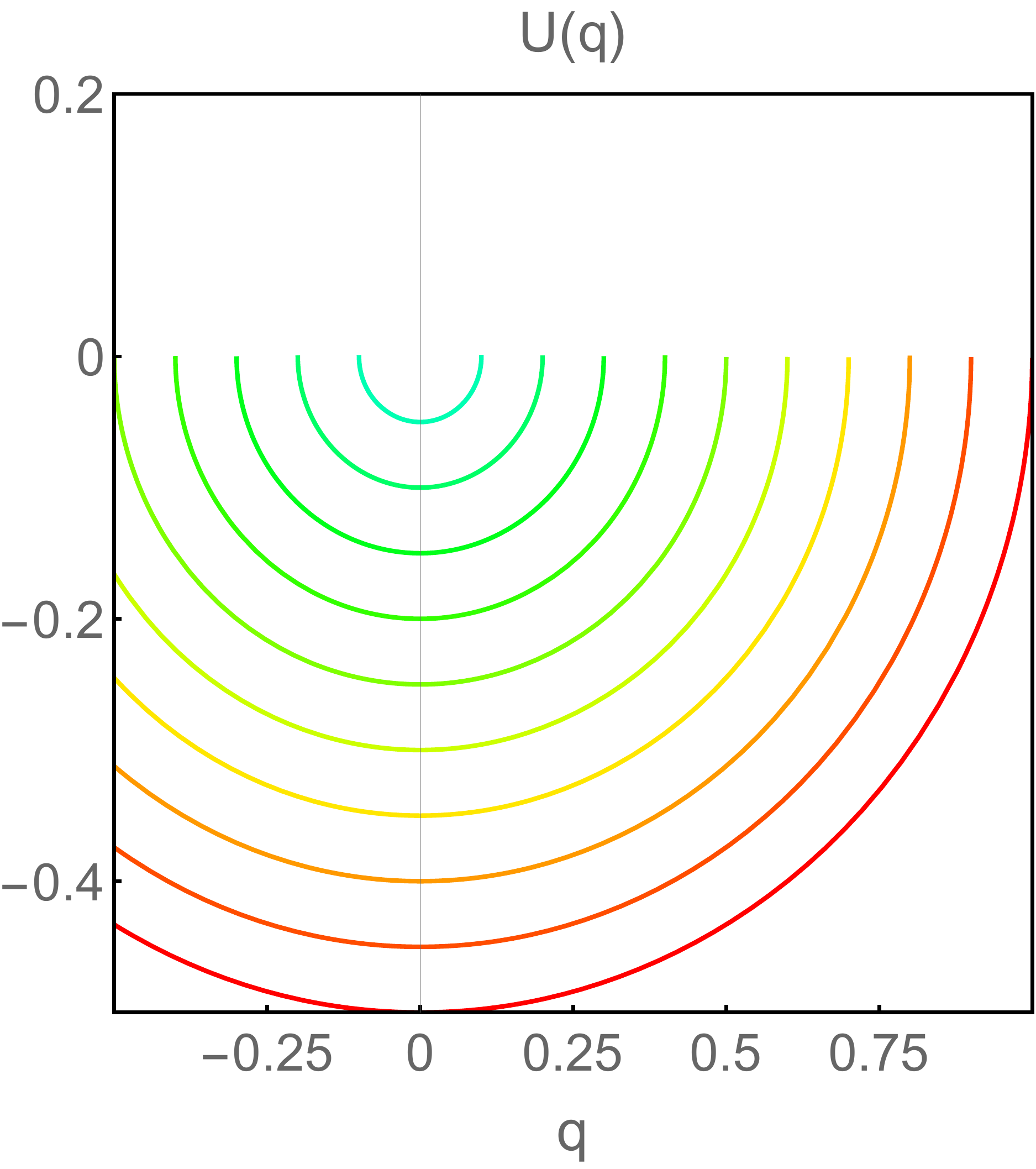} \includegraphics[width=0.32\columnwidth]{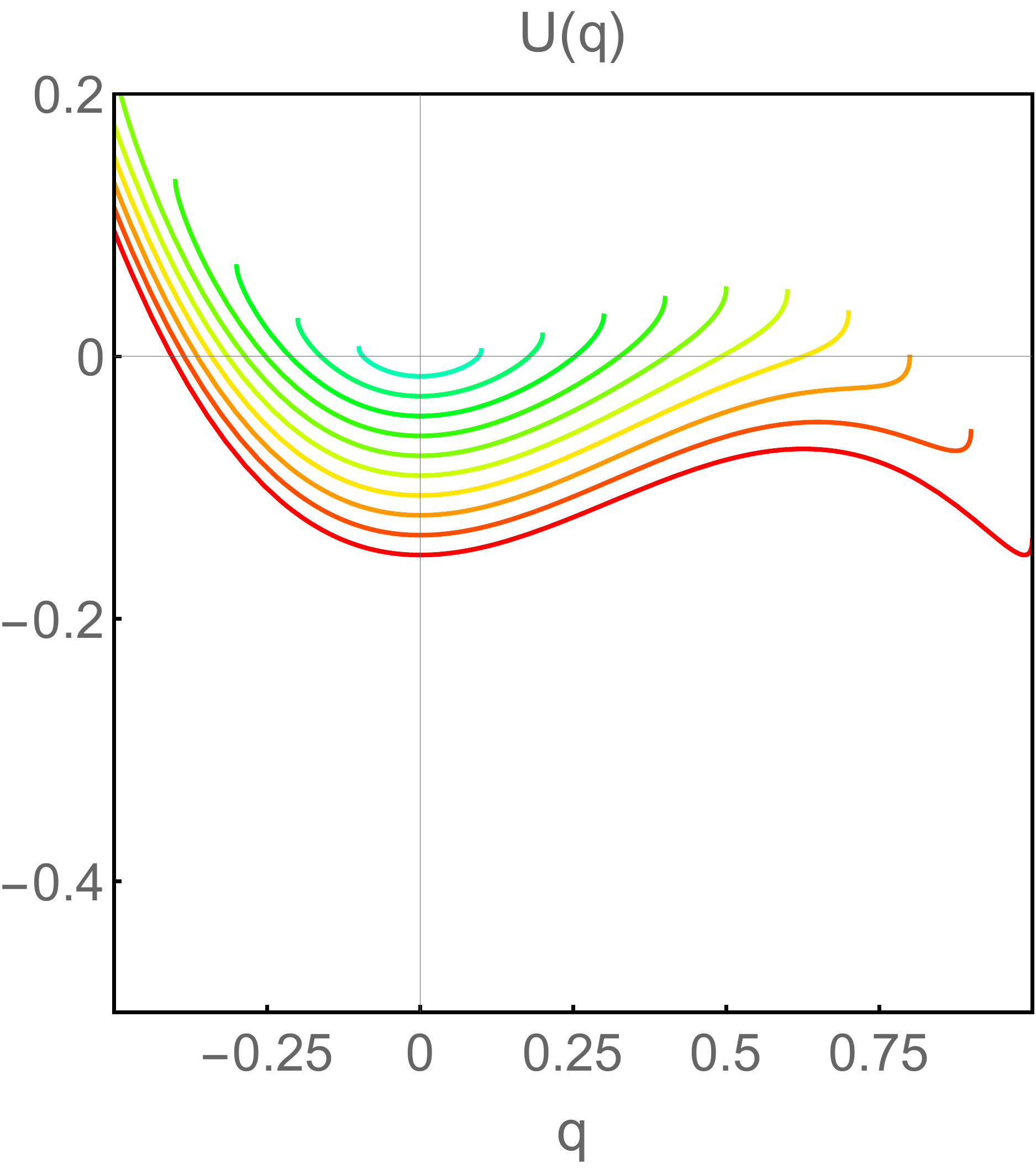}
\includegraphics[width=0.32\columnwidth]{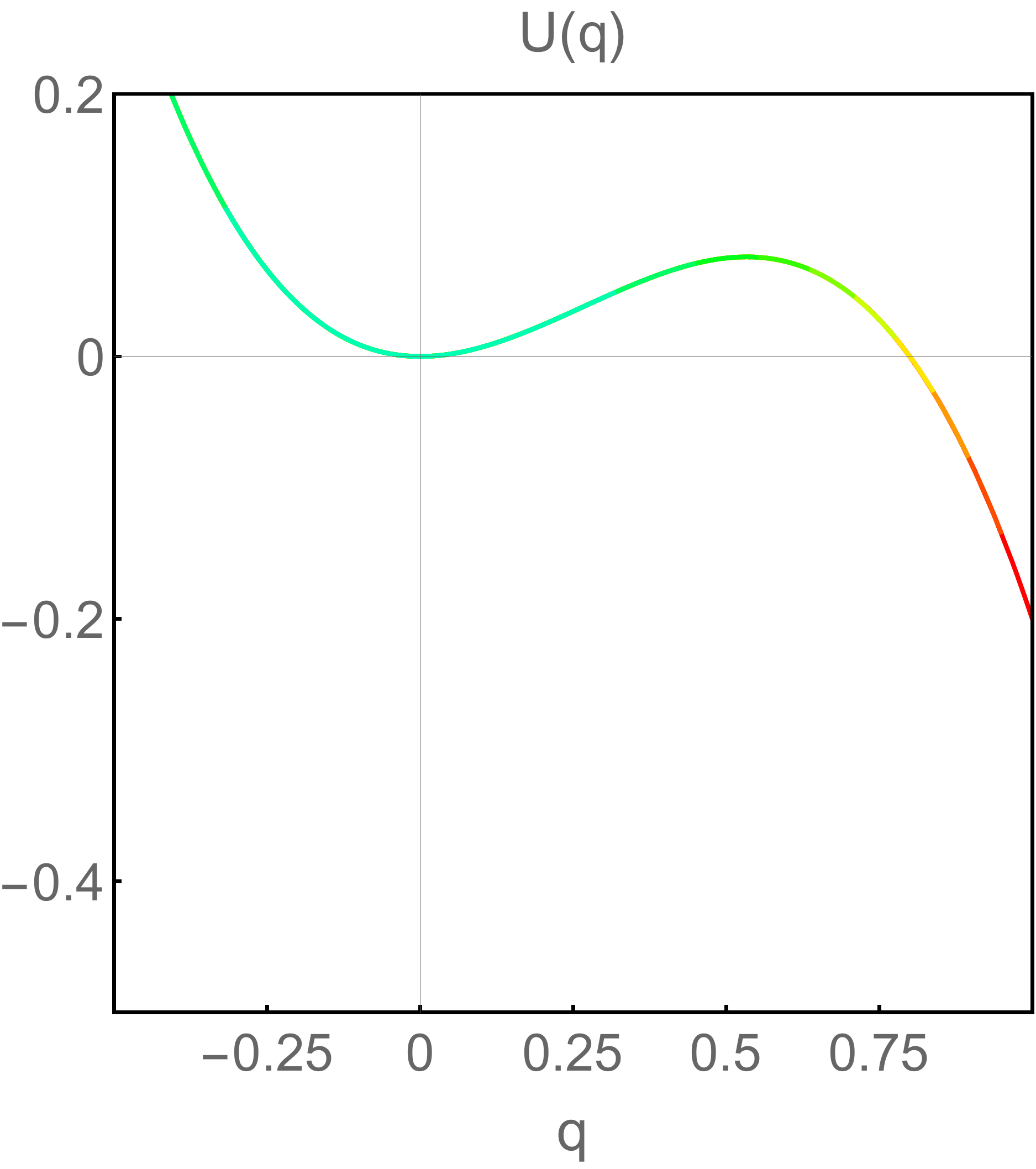}

\includegraphics[width=0.32\columnwidth]{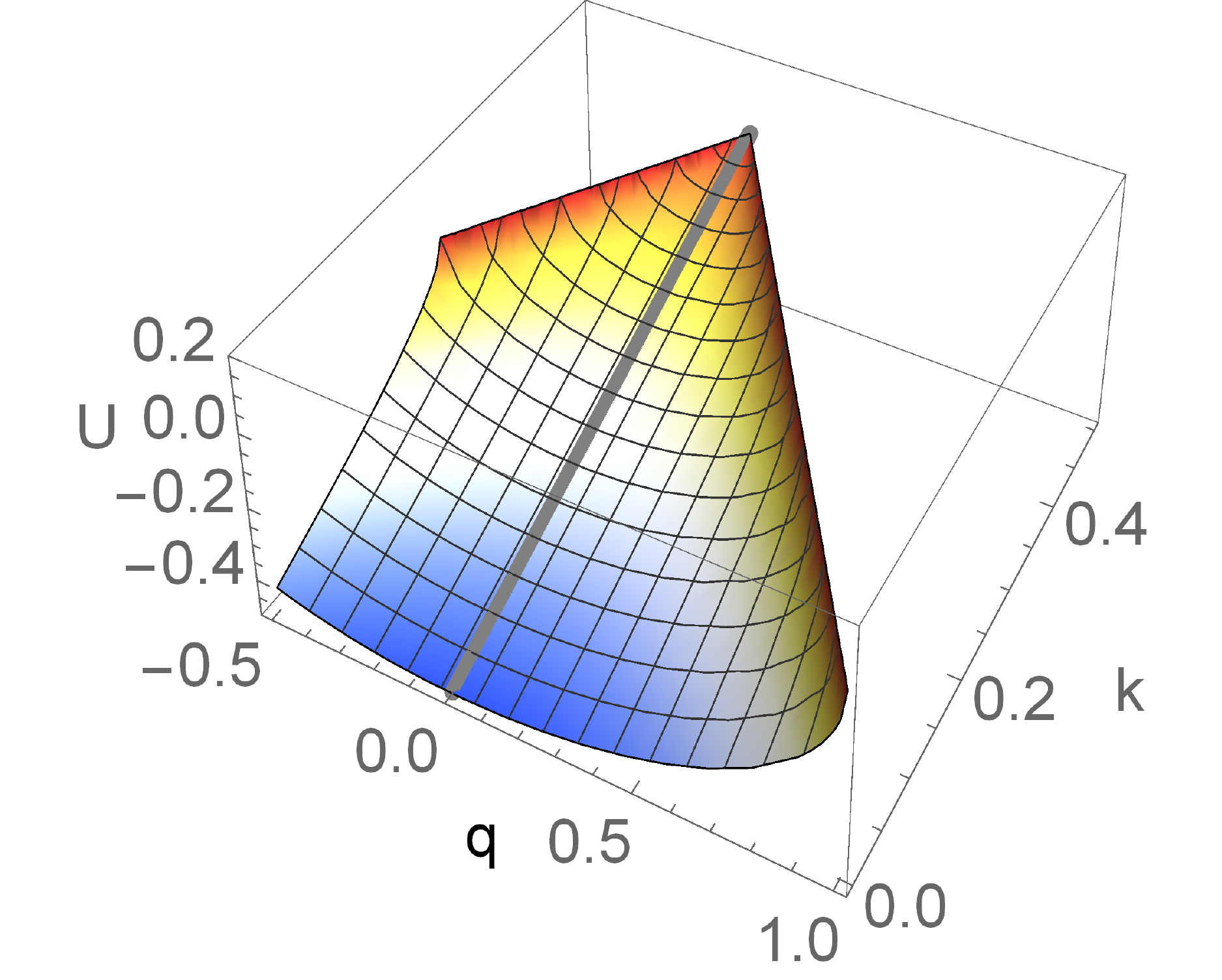} \includegraphics[width=0.32\columnwidth]{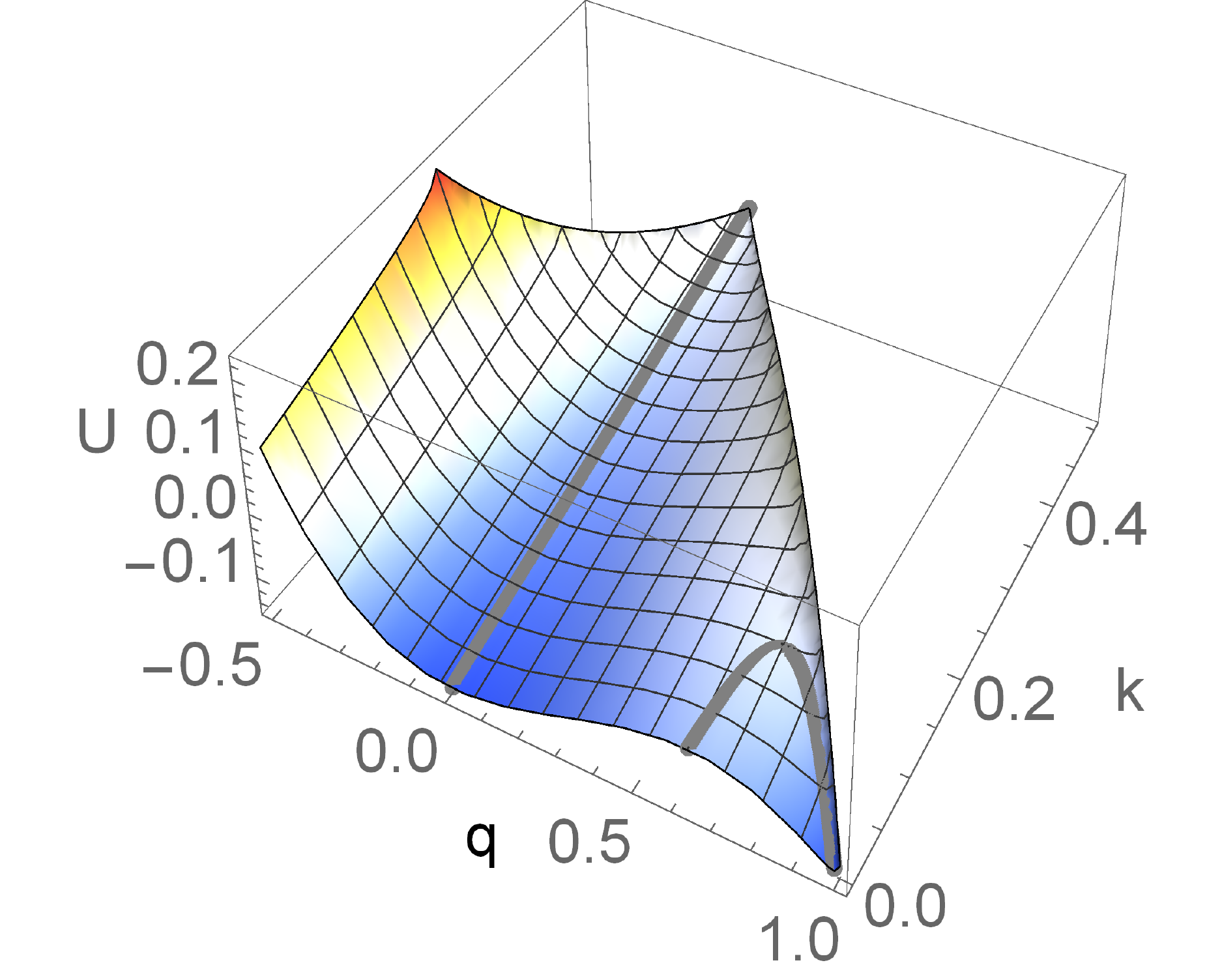}
\includegraphics[width=0.32\columnwidth]{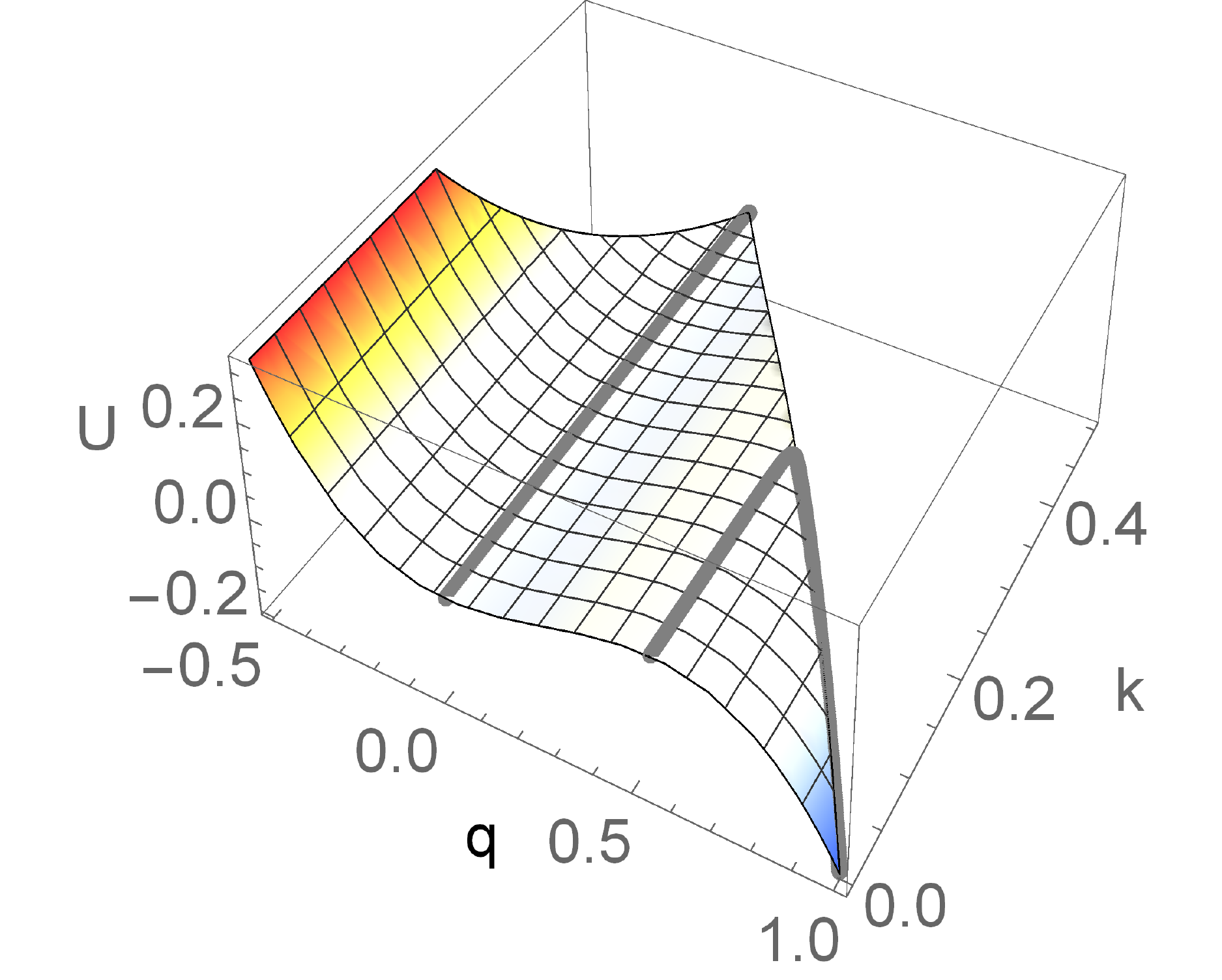}

\protect\caption{Effective classical potential is shown for $s=0$, pure transverse
field case (left), $s=s_{QPT}\approx0.698$, the zero temperature
quantum phase transition point (center), and $s=1$, pure classical
potential (right). Different lines (bottom to top on all plots) correspond
to different values of $k=0,...,0.5$ with equal intervals. Note that
in the absence of the transverse field the states with different values
of $k$ are degenerate. Here $q_{min}=0$ and $q_{max}\approx0.467$.}

\label{fig:U(q)}
\end{figure}

The dynamics of the total spin can be described using a systematic
quasiclassical WKB expansion~\cite{BraunRMP}. In a spin model this
expansion is performed in terms of the small parameter $\varepsilon\equiv2/N\ll1$
which is an analog of $\hbar$ in the textbook WKB approach~\cite{LL3}.
We consider a wave function in the form,

\[
\Psi=e^{i\frac{1}{\varepsilon}\Phi(q)},
\]
and expand it, $\Phi(q)\approx\Phi_{0}+\varepsilon\Phi_{1}+\varepsilon^{2}\Phi_{2}+O\left(\varepsilon^{2}\right)$.
 We can further expand the coefficients $\Phi_{i}$ for a small shift
of the argument from $q$ to $q\pm\varepsilon$,

\begin{eqnarray*}
 & \Psi_{M\pm1}=\Psi_{M}e^{\pm i\dot{\Phi}_{0}}\left(1+i\frac{\varepsilon}{2}\ddot{\Phi}_{0}\pm i\varepsilon\dot{\Phi}_{1}\right)
\end{eqnarray*}
where $\dot{O}\equiv\frac{dO}{dq}$. Substituting this expansion into
the Schroedinger equation we obtain,

\begin{eqnarray*}
 & \mathcal{H}\Psi(q)\approx e(q)\Psi(q),
\end{eqnarray*}
where in the main order in $\varepsilon$ the Hamiltonian is diagonal
and reads, 

\begin{eqnarray}
 & e(q)\approx sf(q)-\frac{1}{2}(1-s)\sqrt{(1-2k)^{2}-q^{2}}\cos\dot{\Phi}_{0}.\label{eq:SecularEquation}
\end{eqnarray}
Note that the function $p\equiv\dot{\Phi}_{0}$ is precisely the canonical
momentum conjugate of the coordinate $q$, $p\rightarrow-i\frac{d}{dq}$.
In other words the second term above is a quantum kinetic energy which
for $p\ll1$ corresponds to a particle with a position dependent mass
$m^{-1}\equiv\frac{1}{2}(1-s)\sqrt{(1-2k)^{2}-q^{2}}$ moving in an
effective classical potential, 

\begin{equation}
U(k,q)=sf(q)-\frac{1}{2}(1-s)\sqrt{(1-2k)^{2}-q^{2}}.\label{eq:EffectivePotential}
\end{equation}
Note that the mass of the effective quantum particle is position dependent,
it increases with increasing $q$ and diverges as $q\rightarrow1$
which affects the efficiency of quantum tunneling into the states
near $q=1$. The effective potential Eq.~(\ref{eq:EffectivePotential})
for different values of $s$ is shown in Fig.~\ref{fig:U(q)}. In
the course of the QA algorithm $s=0\rightarrow s=1$ the effective
potential deforms from a squareroot parabola corresponding to the
ground state $(q,k)=(0,0)$ with the maximal spin polarization along
the transverse field direction at $s\rightarrow0$, see left panel
in Fig~\ref{fig:U(q)}, and the classical potential corresponding
to the ground state $(q,k)=(1,0)$ fully polarized along the axis
of quantization $s\rightarrow1$. Note the initial and final states
of the algorithm are characterized by exponentially small overlap,
see Appendix~\ref{sec:Eigenstates-overlap}.

States with small kinetic energy are confined to one of the potential
wells, centered around the two minima $q_{min}^{(L)}(s)<q_{min}^{(R)}(s)$
of the effective potential $U(k,q)$, which in the course of the evolution
as $s\rightarrow1$ approach the metastable $q_{min}^{(L)}\rightarrow q_{min}$
and ground state $q_{min}^{(R)}\rightarrow1$ of the classical model,
respectively. Confinement of the states is determined by the condition
of vanishing classical velocity, $v(k,q,E)\equiv\frac{\partial\mathcal{H}}{\partial p}=0$,
which gives,
\begin{equation}
-1\leq2\frac{sf(q)-E}{(1-s)\sqrt{(1-2k)^{2}-q^{2}}}\leq1.\label{eq:ClassicalRegion}
\end{equation}
The solution of this equation gives the location of the turning points
$q_{TP}^{(1)}(k,E)$ and $q_{TP}^{(2)}(k,E)$ which limit the classically
allowed region. Inverting the secular equation, Eq.~(\ref{eq:SecularEquation})
for a given energy $E$ we write the wave function in the main order
in $\varepsilon$, 
\begin{equation}
\Psi\sim\exp\left(\tfrac{i}{\varepsilon}\int_{q_{TP}^{(1)}}^{q}dqp\right),\label{eq:WaveFunc}
\end{equation}
with

\begin{equation}
p=\arccos2\frac{sf(q)-E}{(1-s)\sqrt{(1-2k)^{2}-q^{2}}},\label{eq:ConfinedWFPhase}
\end{equation}
when the condition Eq.~(\ref{eq:ClassicalRegion}) is satisfied,
and

\begin{equation}
p=i\mathrm{arccosh}2\frac{sf(q)-E}{(1-s)\sqrt{(1-2k)^{2}-q^{2}}},\label{eq:WFExpTail}
\end{equation}
otherwise. The latter expression corresponds to the exponentially
decaying tail of the wave function extending beyond the classically
allowed region into the potential barrier.   

In the course of the evolution, $s:\,0\rightarrow1$, there is a point
of zero temperature discontinuous quantum phase transition, $s=s_{QPT}$,
at which the minimal energies in the two wells, left $E_{L}$ at $q\sim q_{min}$
and right $E_{R}$ at $q\sim1$, are equal to each other. In the course
of the QA algorithm at finite temperature this transition occurs at
a different transverse field strength $s=s_{PT}(\beta)\geq s_{QPT}$.
The phase transition point can be found from the condition of equal
occupation of the two potential wells $\mathcal{P}_{L}=\mathcal{P}_{R}$
including the entropy of the states. In the large $N$ limit and at
low temperatures $\beta\sim O(N^{0})$ we can approximate the occupation
number $\mathcal{P}_{L}=\sum_{E,k}\exp\left(-\mathcal{F}-\log Z\right)\approx\exp\left(-\mathcal{F}_{L}-\log Z\right)$
by a single dominant term corresponding to a minimum of the free energy.
We write the local minimum condition in a potential well as $\frac{\partial\mathcal{F}}{\partial k}=0$
to obtain,

\begin{equation}
-\beta\frac{\partial E}{\partial k}=-\frac{\partial Q_{k}}{\partial k}\approx\ln\frac{k}{1-k}.\label{eq:OptimizeZ}
\end{equation}
This determines the $k_{min}$ corresponding to the minimum of the
free energy. Whereas the optimal energy (principal quantum number)
with a given $k$ corresponds to the minimum of the potential $U(k_{min},q)$.
We neglect the quantization of levels due to finite $N$ for the purpose
of this calculation. At $s>s_{PT}(\beta)$ the left potential well
is separated from the ground state by a potential barrier with the
shape determined by the parameters $q_{min}$ and $q_{max}$ and overall
scaling $\sim N$. At low temperatures the relaxation in this model
is therefore determined by the rate of transitions between the wells.

In a closed quantum system in the absence of the thermal bath multiple
tunneling effects between the wells need to be taken into account
which corresponds to the ground state formed by a coherent superposition
of the states in the two potential wells, i.e. a Schroedinger cat
state. In a large system the level splitting $\Delta$ corresponding
to such superposition is exponentially small (in the system size $N$),
and therefore such coherent dynamics is quickly suppressed by small
perturbations, such as the hardware noise. This results in overdamped
dynamics characterized by fast intra-well relaxation towards thermal
occupation~\cite{GreenCrowley} reflected in the level width $W\gg\Delta$
and exponentially rare incoherent tunneling effects with the rate
$\sim\Delta^{2}$. On the other hand we neglect the effect of noise
on the tunneling event itself since tunneling is a fast process occurring
on the timescale $1/\omega$, where $\omega$ is the frequency determined
by the curvature of the potential. We assume therefore that $\omega\gg W\gg\Delta$.
We are interested only in the exponential scaling of the transition
rates in this paper ignoring the renormalization of the preexponential
factor that may be substantial in the regime of strong coupling to
the environment. Note that the overdamped dynamics and thermalization
are expected even in the absence of the coupling to a thermal bath,
it can be introduced by a weak disorder in the spin-spin interactions
$\delta H=\sum\varepsilon_{ij}\hat{\sigma}_{i}^{z}\hat{\sigma}_{j}^{z}$
or even weak random transverse field~\cite{Huse}. 

In the regime of the overdamped dynamics at $s<s_{PT}(\beta)$ the
ground state corresponds to $q=q_{min}^{(L)}(s)$. As the system goes
past the phase transition with growing $s>s_{PT}(\beta)$ this state
becomes metastable. The average transition rate $w\ll W$ across
the barrier in presence of fast the intra-well relaxation $W$ can
be obtained by calculating the total current escaping the metastable
well~\cite{Chudnovsky98,ChudnovskyBook},
\begin{quotation}
\begin{equation}
w\propto\frac{1}{Z}\sum_{k,E}w(k,E)e^{-\beta E+Q_{k}}.\label{eq:TransitionRateT}
\end{equation}

\end{quotation}
Eq.~(\ref{eq:TransitionRateT}) is a thermal average, weighted with
the usual Boltzmann factor $e^{-\beta E+Q_{k}}$ including the entropy
$Q_{k}$, of $w(k,E)\sim e^{-S_{WKB}(k,E)}$, the incoherent tunneling
amplitude through the barrier of a state with a given energy $E$
and total spin parameter $k$. The so called reduced action $S_{WKB}(k,E)$
can be obtained by matching the quasiclassical wave functions across
the barrier region or following the analytical continuation procedure~\cite{LL3},

\begin{eqnarray*}
 & S_{WKB}(k,E)=-2\int_{q_{L}}^{q_{R}}dqp(q)\\
 & =-N\int_{q_{L}}^{q_{R}}dq\mathrm{arccosh}\frac{2\left(sf(q)-E\right)}{(1-s)\sqrt{(1-2k)^{2}-q^{2}}}.
\end{eqnarray*}

The sum Eq.~(\ref{eq:TransitionRateT}) can be approximated by its
largest term, the rest being exponentially smaller,
\begin{quotation}
\begin{equation}
w\sim\frac{1}{Z}\sum_{k,E}w(k,E)e^{-\beta E+Q_{k}}\approx e^{-S_{opt}}.\label{eq:SteepestDescent}
\end{equation}

\end{quotation}
Where the largest term is found by minimizng the action,
\begin{equation}
S_{opt}=S_{WKB}+\beta E-Q_{k}+\ln Z\label{eq:OptimalAction}
\end{equation}
with respect to $k$ and $E$,

\begin{eqnarray}
 & T(E)\equiv-\frac{\partial S_{WKB}}{\partial E}=\beta\label{eq:OptimalPeriod}\\
 & -\frac{\partial S_{WKB}}{\partial k}-\beta\frac{\partial E}{\partial k}=-\frac{\partial Q_{k}}{\partial k}\approx\ln\frac{k}{1-k}.\label{eq:OptimalSpin}
\end{eqnarray}
where in Eq.~(\ref{eq:OptimalPeriod}) fixed $k$ is assumed. Since
$Q_{k}$ is independent of the energy level $E$ the conditions on
the optimal tunneling parameters separate into the standard condition
Eq.~(\ref{eq:OptimalPeriod}) requiring the period of motion in the
inverted potential $T(E)$ to match the inverse temperature $\beta$~\cite{ChudnovskyBook},
and the condition Eq.~(\ref{eq:OptimalSpin}) due to the entropy
of states dependent on $k$, which introduces novel physics in the
dynamics of this model. Eqs.~(\ref{eq:OptimalAction},~\ref{eq:OptimalPeriod},~\ref{eq:OptimalSpin})
need to be supplemented with conditions ensuring that the energy and
total spin $k$ are conserved in the tunneling event (we emphasize
that we neglect here the effect of the thermal bath during the tunneling
event). The tunneling from the metastable well has to be at an energy
and spin values, $E$ and $k$, at which a state exists in the ground
state well, i.e. $E\geq\mathrm{min}\left\{ E_{R}(k)\right\} $, which
is not always satisfied in the system with large entropy of states,
i.e. $\mathcal{F}_{L}<\mathcal{F}_{R}$ does not necessarily imply
$E_{L}<E_{R}$.

\begin{figure}
\includegraphics[width=0.49\columnwidth]{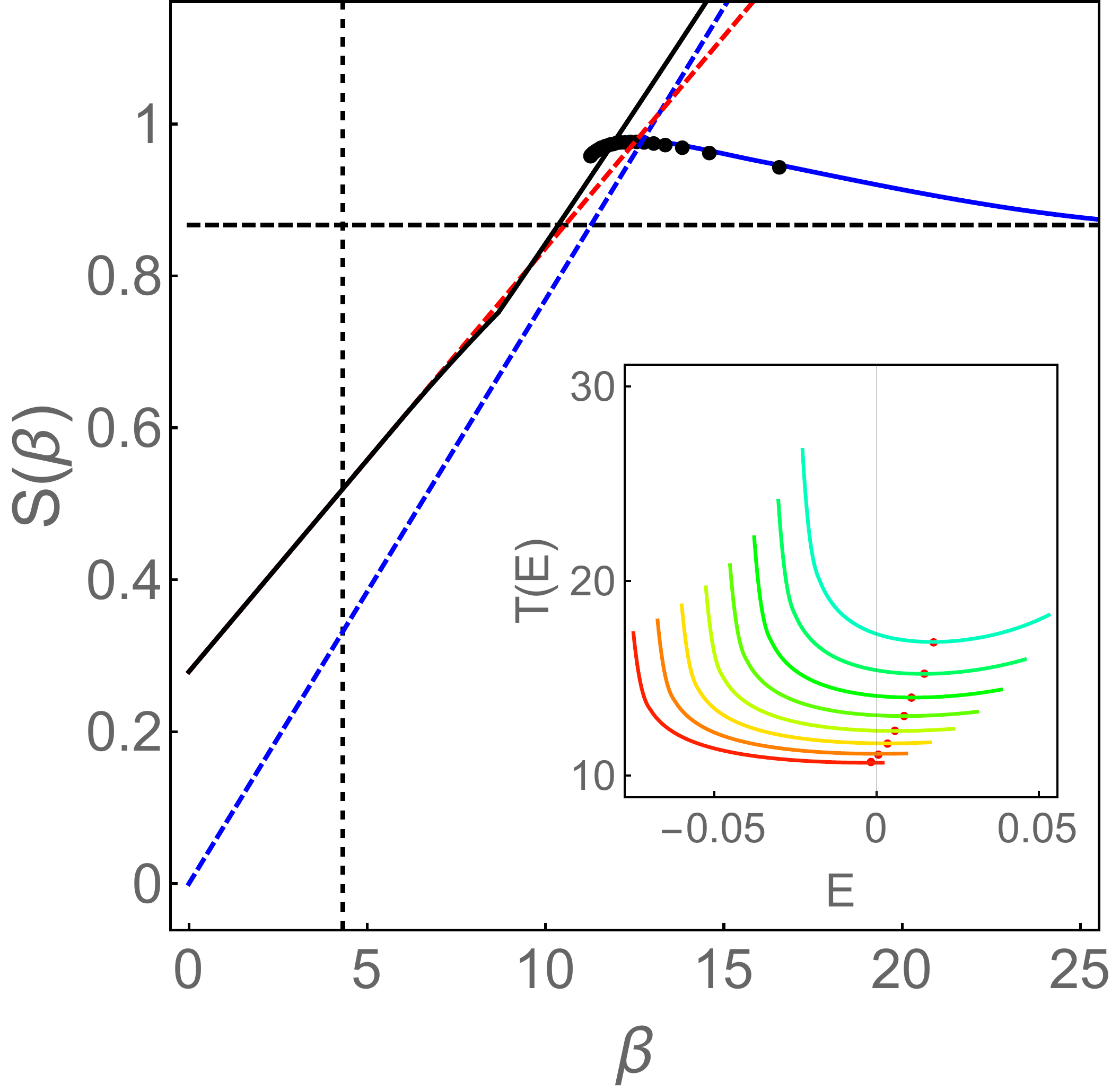} \includegraphics[width=0.49\columnwidth]{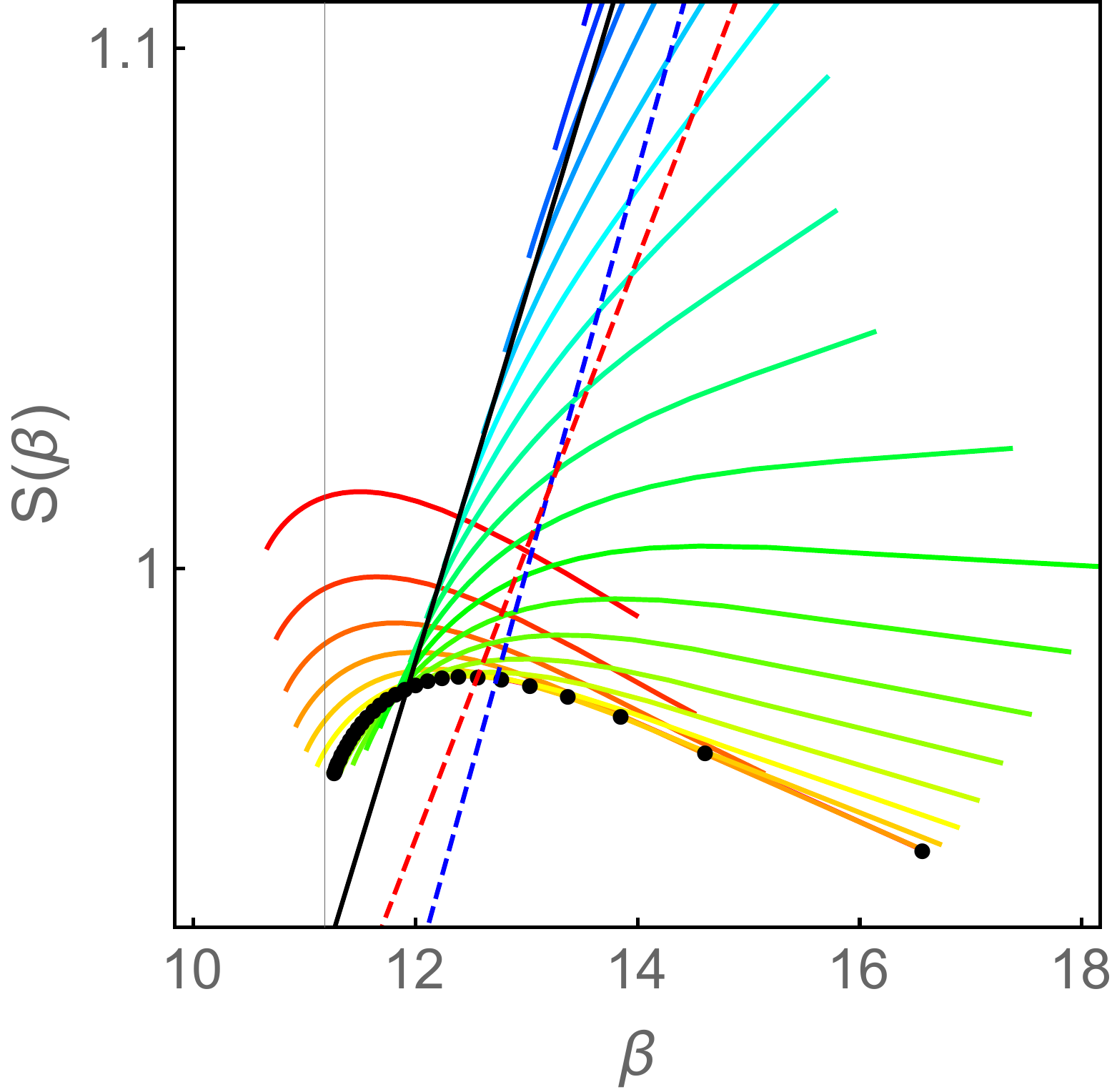}

\protect\caption{Tunneling action as a function of inverse temperature $\beta$. For
$q_{min}=0,\, q_{max}=0.533$ at $s=0.85>s_{QPT}$. \emph{Left:} Black
dots, and the solid blue line fit, correspond to optimal finite temperature
quantum tunneling action. Horizontal dotted line is the $\beta\rightarrow\infty$
limit of the incoherent tunneling action corresponding to the quantum
mechanical tunneling from the lowest level of the metastable well
corresponding to $k=0$. Solid black line corresponds to the optimal
action of the classical Glauber dynamics. Blue and red dashed lines
show the classical over-the-barrier escape action at $k=0$ and along
the line $q=1-2k$ with the potential maximum at $k_{\ast}\approx0.152$,
respectively. The latter, $k_{\ast}$, is the point of inflection,
i.e. the point where the minimum of the right potential well merges
with the maximum at the barrier top, see Fig.~\ref{fig:U(q)}. Vertical
dashed line corresponds to the temperature driven phase transition
point $\beta_{PT}\approx4.32$ corresponding to equal occupation of
the two potential wells.\emph{ Inset:} $T(E)=-\partial S/\partial E$
at fixed $k$ which corresponds to the period of quantum mechanical
tunneling trajectory in the imaginary time representation. Dominant
contribution comes from tunneling at the energy determined from $T(E)=\beta$
for $\beta>T_{min}$, at $\beta<T_{min}$ the transition rate is dominated
by the over-the-barrier escape. Different lines bottom to top correspond
to different values of $s$ at fixed $k=0$. Red dots correspond to
$T_{min}$. Note that the minimum $E_{min}:\, T(E_{min})=T_{min}$
does not correspond to the highest tunneling energy. This means that
the transition from quantum tunneling regime to over-the-barrier escape
has discontinuous 1st order character. \emph{Right: }Quantum-classical
transition region on a larger scale. Black dots correspond to tunneling
action. Different lines show quantum mechanical tunneling action for
fixed $k$, colors correspond to growing $0\leq k\leq0.152$ (red
to blue). Quantum tunneling process conserves energy and total spin
value $k$. At $k>k_{LR}$ the point where $E_{L}(k_{LR})=E_{R}(k_{LR})$
quantum mechanical tunneling process requires the state in the metastable
well to be at an energy $E\geq E_{R}(k)$ which comes with an additional
thermal excitation cost $\beta\left(E_{R}(k)-E_{L}(k)\right)$ in
the tunneling action. Therefore for growing $k$ the tunneling action
$S(\beta)$ resembles linear classical dependence. Solid black and
dashed red and blue lines are the same as in the left figure.}

\label{fig:S(Beta)}
\end{figure}

Eq.~(\ref{eq:OptimalPeriod}) has a solution in a range of energies
$E$ such that $T_{min}\leq T(E)<\infty$, see inset in Fig.~\ref{fig:S(Beta)}
left.  In the case of $\beta>T_{min}$ the quantum tunneling process
dominates in the sum in Eq.~(\ref{eq:SteepestDescent}). For $\beta<T_{min}$
there are no solutions to Eq.~(\ref{eq:OptimalPeriod}) and therefore
the optimal energy is at the edge of the interval $E=U(q_{max})-U(q_{min})$
corresponding to the height of the barrier. In other words, in this
regime over-the-barrier escape process dominates with $\beta\sim T_{min}$
being the point of a quantum-to-classical phase transition. Note that
the global minimum $E_{min}$ of the function $T(E)$ in the inset
in Fig.~\ref{fig:S(Beta)} does not always correspond to the top
of the barrier, which means that the quantum to classical transition
(in the limit $N\rightarrow\infty$) has a discontinuous 1st order
character~\cite{Owerre20151}. Considering Eq.~(\ref{eq:OptimalSpin})
we look for a solution in the interval $0\leq k\leq k_{\ast}$ where
$k_{\ast}$ is the inflection point of the potential $U(q,k)$ where
the right (the ground state) potential well disappears, see Fig.~\ref{fig:U(q)},
since quantum tunneling conserves the total spin $k$, for it to occur
there must exist states with matching $k$ in the ground state potential
well.  

The result of this optimization procedure is the optimal action $S_{opt}(\beta)$
at a fixed $s$ shown in Fig.~\ref{fig:S(Beta)}. In the vanishing
temperature limit $\beta\rightarrow\infty$, where the effect of entropy
on the occupation of levels is negligible, $S_{opt}(\beta)$ corresponds
to the quantum tunneling from the lowest energy level in the metastable
well corresponding to $k=0$ (horizontal dashed line in Fig.~\ref{fig:S(Beta)}
left). As temperature increases ($\beta$ decreases) the entropy starts
playing a role in dynamics and $S_{opt}(\beta)$ increases up to some
maximum value, see blue solid line in Fig.~\ref{fig:S(Beta)} left,
in contrast with the usual (non-degenerate) case~\cite{ChudnovskyBook}
where quantum tunneling rate increases with increasing temperature
$\beta^{-1}$. This is a result of the entropy providing a high statistical
weight to the state with suboptimal tunneling rate. This behavior
is not entirely generic and originates in the model of Eqs.~(\ref{eq:Hamiltonian})
and (\ref{eq:CubicPotential}) because the transverse field lifts
the degeneracy of the metastable state favoring the non-degenerate
state with $k=0$, see Fig.~\ref{fig:U(q)}. Whereas the entropy
favors the state with $k=1/2$ (at $\beta\rightarrow0$) and therefore
with increasing temperature the state with the highest statistical
weight will correspond to a finite $1/2\geq k>0$, given by the solution
of Eq.~(\ref{eq:OptimizeZ}). At the same time the potential barrier
increases with $k$ and therefore the quantum tunneling becomes less
and less efficient, resulting in increasing $S_{opt}(\beta)$ with
decreasing $\beta$ up to the regime of the transition into the classical
escape at high temperatures $\beta\ll T_{min}$. The classical over-the-barrier
excitation is described by the thermal excitation rate $w_{cl}\sim Z^{-1}\exp(-\beta E+Q_{k})$.
Note that the classical process is driven by the Glauber dynamics
of the spins due to the effect of the thermal bath. This process does
not conserve the total spin $k$ and therefore the optimal classical
trajectory is determined by the saddle point of the free energy (including
the entropy) in the 2D space of $(k,q)$. Entropy provides an additional
cost reducing the transition rate which is reflected in the finite
offset of the dependence of the classical transition rate on $\beta$
as $\beta\rightarrow0$, see black solid line in Fig.~\ref{fig:S(Beta)}
Left, see Appendix~\ref{sec:Simulated-Annealing} for more details.
Blue and red dashed lines in Fig.~\ref{fig:S(Beta)} indicate the
linear in $\beta$ dependence of the energy cost of the over the barrier
excitation at $k=0$ and along the line of $q=1-2k$ with the potential
maximum at $k\approx k_{\ast}$ (the latter is the inflection point
of the potential $U(k,q)$), respectively, which correspond to the
low temperature potential energy dominated regime and the high temperature
entropy dominated regime~\footnote{Note that in the case of the Grover's unstructured search problem a completely flat potential,$U(k,q)=0$ everywhere except $U(0,1)=1$, the offset of the classical action in Fig.~\ref{fig:S(Beta)} corresponds to the classical complexity (computation time) of the Grover's problem with respect to simulated annealing, $2^N$. Whereas in presence of the potential profile $U(k,q)\neq0$ which provides structure to the problem the complexity may be lower.}.

The steepest descent approximation Eq.~(\ref{eq:SteepestDescent})
is applicable as long as the preexponential factors in the sum are
non-divergent which is true away from $\beta\sim T_{min}$ a phase
transition point from the quantum tunneling regime to the classical
over-the-barrier escape regime~\cite{ChudnovskyBook}. On general
grounds we expect the action to be continuous even in the case where
this quantum-to-classical transition is of the 1st order, with the
discontinuity occurring in the derivative of the action~\cite{Owerre20151}.
Therefore we expect Fig.~\ref{fig:S(Beta)} to provide a qualitatively
correct dependence of $S(\beta)$ in the whole range of inverse temperatures
$\beta$.

\section{Quantum annealing computation time\label{sec:Computation-time}}

In the course of the QA algorithm the transverse field parameter $s$
is varied from $s=0$ to $s=1$ with a fixed rate $v$ at a fixed
inverse temperature $\beta$. The goal is to find the ground state
with probability approaching $\sim1$, allowing for repeated runs
of the algorithm. Here we assume that finding any state within the
ground state potential well is sufficient to find the solution (as
local search could allow identifying the lowest energy state within
the well). Given the probability $\mathcal{P}_{GS}$ of finding the
ground state after a single run of duration $v^{-1}$, the number
of runs needed to achieve this goal is $\mathcal{P}_{GS}^{-1}$. The
total computation time is therefore given by, 

\[
\tau\sim v^{-1}\times\mathcal{P}_{GS}^{-1}.
\]
The quantum mechanical tunneling rate vanishes as $s\rightarrow1$
(at very low temperatures of interest here the over-the-barrier transition
rate is also weak). Therefore there exists a point $s=s_{F}$ in the
course of the sweep of the transverse field where the relaxation time
$\sim w^{-1}(s_{F})$ required to achieve thermal distribution becomes
longer than the length of the algorithm, $w^{-1}(s_{F})\sim v^{-1}$.
In other words the computation will be finished before the thermal
equilibrium is reached. The system effectively freezes the values
of the occupation numbers of the left $\mathcal{P}_{L}$ and right
$\mathcal{P}_{R}$ wells at the last point (in the course of the algorithm)
where the inter-well transition process was still fast enough (the
intra-well relaxation may still be efficient at $s>s_{F}$). We call
this the freezing point. The computation time of the algorithm (its
exponential scaling) is therefore determined by the occupation probability
$\mathcal{P}_{GS}=\mathcal{P}_{R}(s_{F}),$ and the equilibration
time $w^{-1}(s_{F})$ at the freezing point,

\begin{equation}
\xi\equiv\frac{1}{N}\log\tau\approx\frac{1}{N}\log\left[e^{NS_{opt}}\left(1+e^{N(\mathcal{\mathcal{F}}_{R}-\mathcal{F}_{L})}\right)\right],\label{eq:QAComputationTime}
\end{equation}
where $S_{opt}$ is given by Eq.~(\ref{eq:OptimalAction}), with
$s=s_{F}.$ The optimal computation time of the quantum annealing
is found by minimizing with respect to the location of $s_{F}$ and
the temperature $\beta$ at which the computation is performed. The
point of the phase transition, $\beta=\beta_{PT}$ and $s_{F}=s_{PT}$,
respectively, defined by, $\mathcal{\mathcal{F}}_{R}-\mathcal{F}_{L}=0$
separates two scaling regimes of the computation time, 

\begin{eqnarray}
 & \xi\approx\left[\begin{array}{cc}
S_{opt}+\mathcal{\mathcal{F}}_{R}-\mathcal{F}_{L}, & \beta<\beta_{PT}(s)\\
S_{opt}, & \beta>\beta_{PT}(s)
\end{array}\right.\label{eq:ComputationTimeScaling}
\end{eqnarray}
The high temperature limit of this expression is given by the entropy
difference between the two wells. This is the limit of local exhaustive
search. This is always a bound on the computation time of the algorithm
with which we need to compare for completeness. We first consider
$\beta<\beta_{PT}$ assuming the free energy of the system demonstrates
two minima $\mathcal{\mathcal{F}}_{R}$ and $\mathcal{F}_{L}$. In
the quantum tunneling regime both $S_{opt}(\beta)$ and $\mathcal{\mathcal{F}}_{R}-\mathcal{F}_{L}$
decrease with growing $\beta$ and therefore $\xi(\beta)$ is monotonously
decreasing as well. In the classical regime $S_{opt}+\mathcal{\mathcal{F}}_{R}-\mathcal{F}_{L}$
is also a monotonous function which depending on competition between
$S_{opt}(\beta)$ and $\mathcal{\mathcal{F}}_{R}-\mathcal{F}_{L}$
can be either increasing, in which case $\beta\rightarrow0$ is the
optimal classical computation regime (i.e. the local exhaustive search
limit), or decreasing towards the critical point $\beta=\beta_{PT}$.
Therefore we need to analyze the performance in the regime $\beta>\beta_{PT}$
and $s>s_{PT}$ and compare it to local exhaustive search. The optimal
computation time in this regime, see Eq.~(\ref{eq:ComputationTimeScaling}),
is determined by the minimum of the transition rate $S_{opt}$ with
respect to $\beta$ and $s$.

Because of the concave dependence of $S_{opt}(\beta)$ on the inverse
temperature, as shown in Fig.~\ref{fig:S(Beta)} left, the minimum
of the action with respect to $\beta$ corresponds to one of the edges
of the inverse temperature interval, i.e. either $\beta=\beta_{PT}$
or $\beta\rightarrow\infty$. The global minimum is therefore the
smallest of $S(\beta_{PT}(s_{F}),s_{F})$ and $S(\infty,s_{F})$.
The minimum with respect to $s_{F}$ (and therefore the global minimum)
is determined by comparing these two functions.

\begin{figure}
\includegraphics[width=0.48\columnwidth]{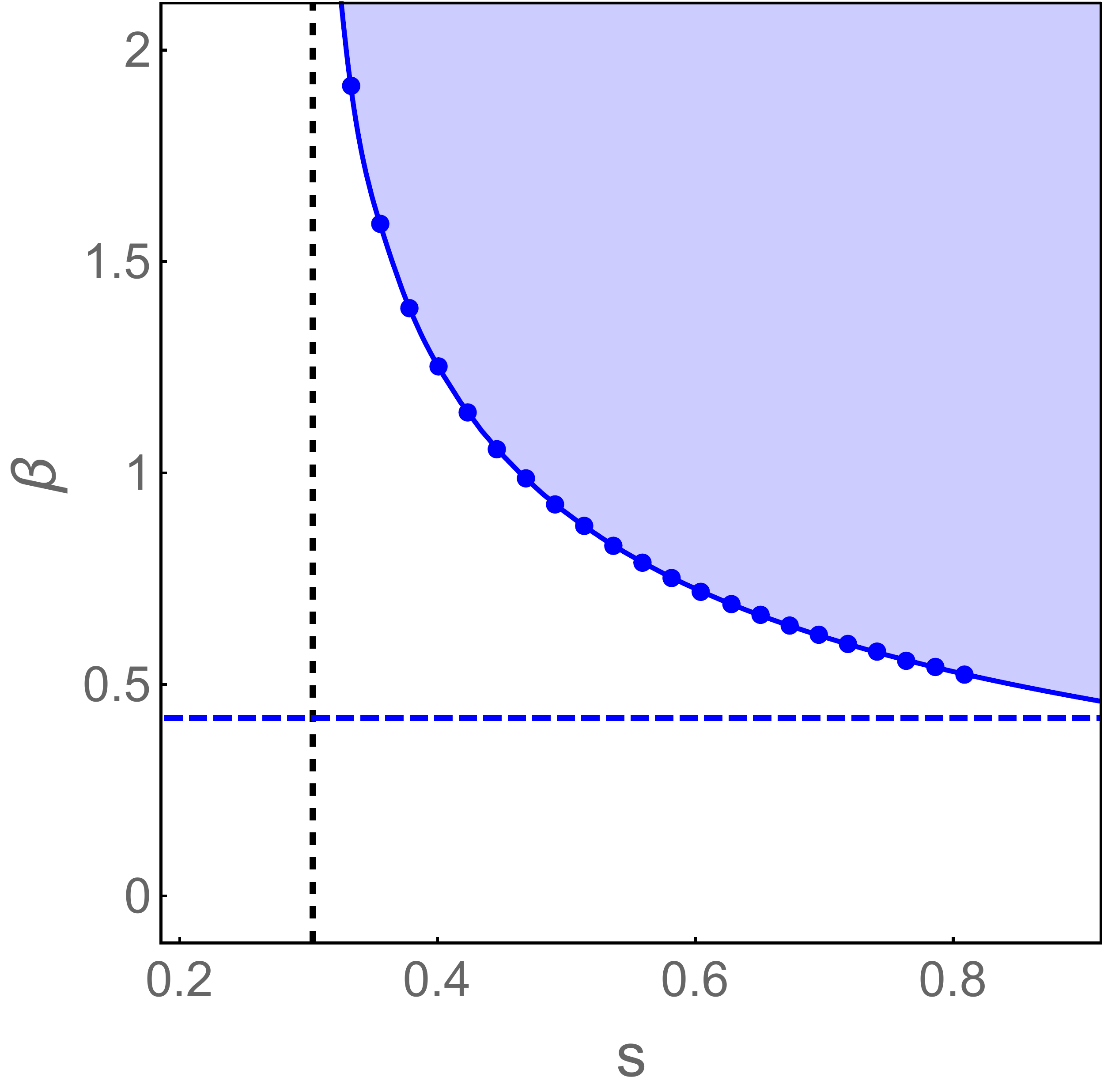}
\includegraphics[width=0.485\columnwidth]{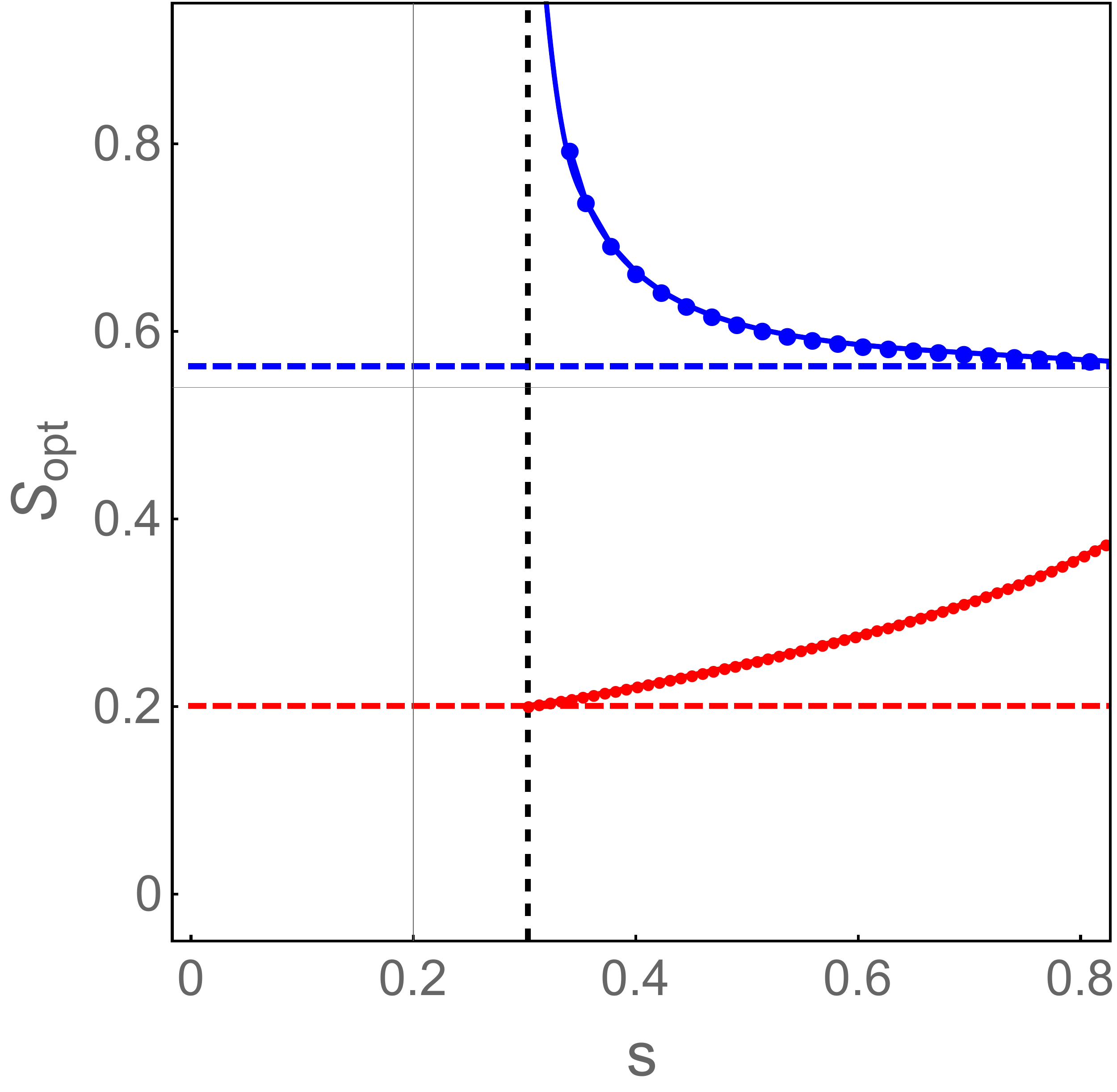}

\protect\caption{\emph{Left:} Inverse critical temperature dependence on transverse
field strength $\beta_{PT}(s)$ for $q_{min}=0.88$, $q_{max}=0.955$.
The horizontal dashed line corresponds to the classical transition
temperature and the vertical dashed line marks the point of the quantum
phase transition. \emph{Right: }Blue line shows optimal over-the-barrier
escape action along the critical line\emph{ }shown in the left panel\emph{.
}Blue points correspond to the same values of $s$ as in the left
figure. Horizontal blue line corresponds to the transition rate at
the critical temperature in the classical model, $s=1$. Vertical
line corresponds to the point of zero temperature quantum phase transition.
Red line corresponds to the vanishing temperature limit of the quantum
tunneling action. Red horizontal dashed line corresponds to the minimum
of the quantum tunneling action at the quantum phase transition point.
The local exhaustive search corresponds to $\xi=\frac{1}{N}\log\tau_{es}=Q_{cl}(q_{min})\approx0.227$,
larger than the quantum annealing time.}

\label{fig:CriticalLineAction}
\end{figure}

\begin{figure}
\includegraphics[width=0.49\columnwidth]{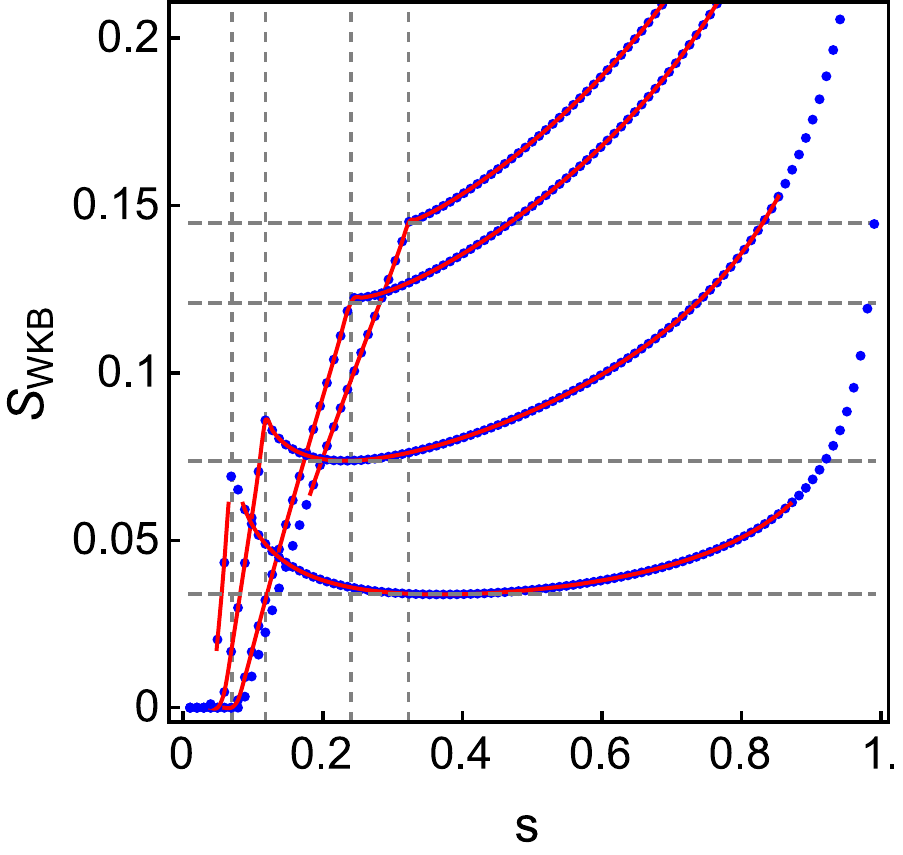} \includegraphics[width=0.48\columnwidth]{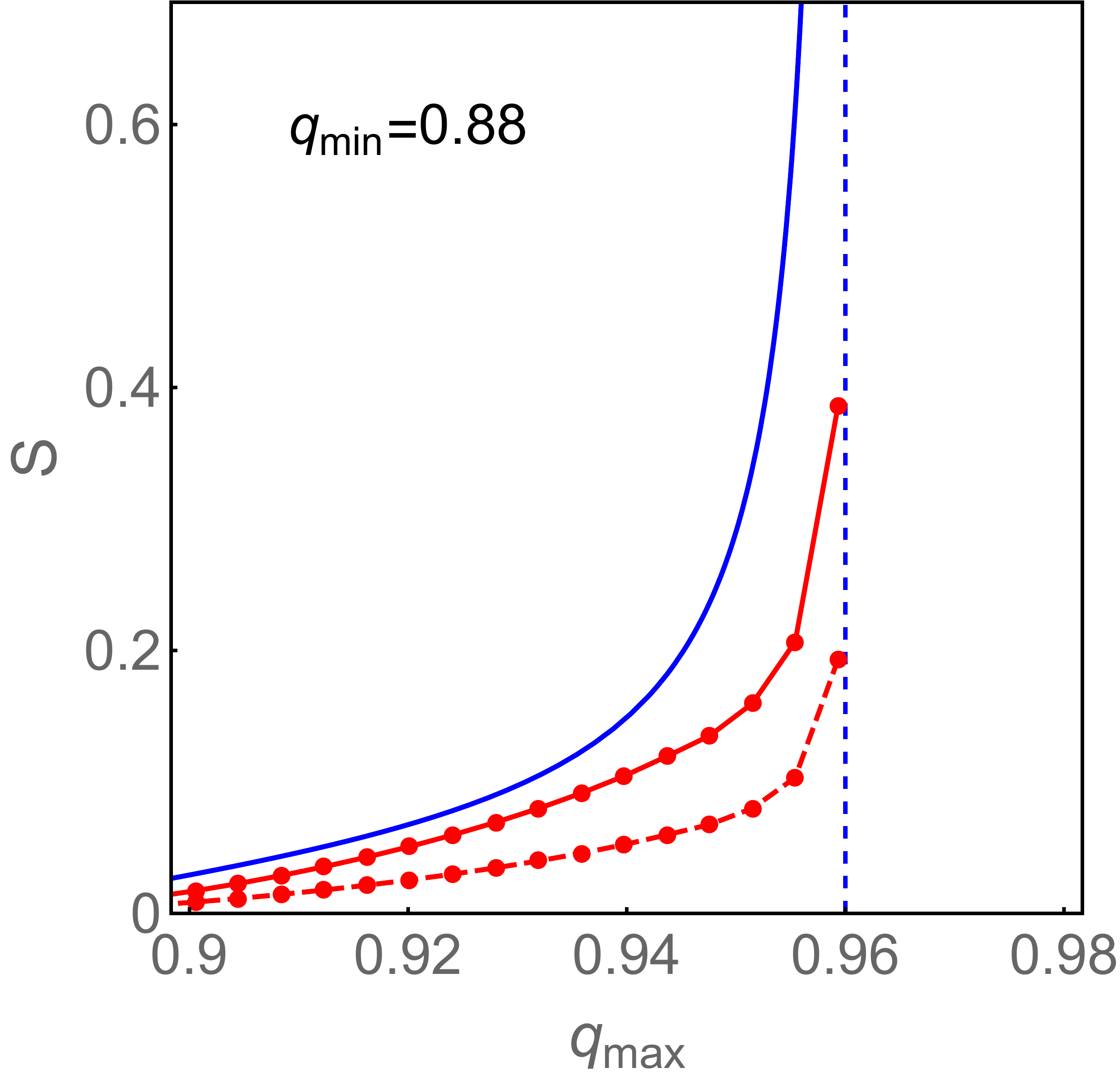}

\protect\caption{\emph{Left:} Quantum mechanical tunneling action at vanishing temperature
in the incoherent regime as a function of the transverse field parameter
$s$. Different curves correspond to different values of $q_{max}=0.929,\,0.946,\,0.958,\,0.961$
with fixed $q_{min}=0.9$. Note that the optimal tunneling rate emphasized
by horizontal dashed lines does not always correspond to the phase
transition point denoted by vertical dashed lines. In other words
the optimal QA schedule in the presence of dissipation is crucially
different from QA in an isolated system. \emph{Right: }Optimal action
for SA (blue) and QA (red) at $q_{min}=0.88$ for different values
of $q_{max}$. The vertical dashed lines corresponds to $q_{min}=\frac{3}{2}q_{max}$
at which point the classical potential has a degenerate ground state
at $q=q_{min}$ and $q=1$. At this point both SA action and QA action
diverge, however QA action diverges logarithmically slow in contrast
to SA action. For the parameters chosen the local exhaustive search
corresponds to $\xi=\frac{1}{N}\log\tau_{es}=Q_{cl}(q_{min})\approx0.227$.}

\label{fig:ZeroTAction}
\end{figure}

A typical critical line is shown in the left panel of Fig.~\ref{fig:CriticalLineAction}.
The inverse critical temperature $\beta_{PT}(s)$, blue solid line
in Fig.~3 left, diverges at the point of the quantum phase transition
in the course of the algorithm, $s=s_{QPT}$ (vertical dashed line
in Fig.~\ref{fig:CriticalLineAction}), and at $s>s_{QPT}$ monotonously
decreases with $s$ approaching the classical transition temperature
at $s=1$ (horizontal dashed line in Fig.~\ref{fig:CriticalLineAction}).
The right panel of Fig.~\ref{fig:CriticalLineAction} shows the optimal
over-the-barrier escape action at the given critical temperature $\beta_{PT}(s)$,
blue (upper) solid line. The blue dots in the left and right panels
correspond to the same values of $s$. The classical action, following
the behavior of the inverse critical temperature diverges at the point
of the quantum phase transition $s=s_{QPT}$ and decreases with $s$
in the course of the algorithm approaching the value corresponding
to the classical model $s=1$, shown as the blue (upper) dashed line.
The latter is the optimum (due to the highest critical temperature)
and therefore defines the optimal computation time of the classical
simulated annealing algorithm, see Appendix for a more details. The
optimum of the quantum action, at vanishing temperature, \textbf{$\beta\rightarrow\infty$},
is given by the tunneling from the bottom of the metastable well.
In this limit the entropy does not affect the occupation of the energy
levels in the course of the algorithm. We assume that the temperature
is still high enough such that there is sufficiently fast intra-well
relaxation. Fig.~\ref{fig:CriticalLineAction} right shows $S_{opt}(\infty,s)$,
red (lower) solid line, which assumes the minimal value at the quantum
phase transition point $s=s_{QPT}$. Note that this is not true for
all the parameter values, instead the minimum of $S_{opt}(\infty,s)$
often occurs at some $s>s_{QPT}$ and its value at this point can
be as much as two times smaller than the value at $s=s_{QPT}$, see
Fig.~\ref{fig:ZeroTAction} left. It is possible to take advantage
of this global minimum only in the course of the open system Quantum
Annealing since the transition corresponding to this rate maximum
occurs not into the ground state level but into one of the excited
states. Therefore such a transition should necessarily be followed
by relaxation within the local domain of attraction. Such local relaxation
is not available in the closed system adiabatic quantum annealing
algorithm.

\begin{figure}
\includegraphics[width=0.3\columnwidth]{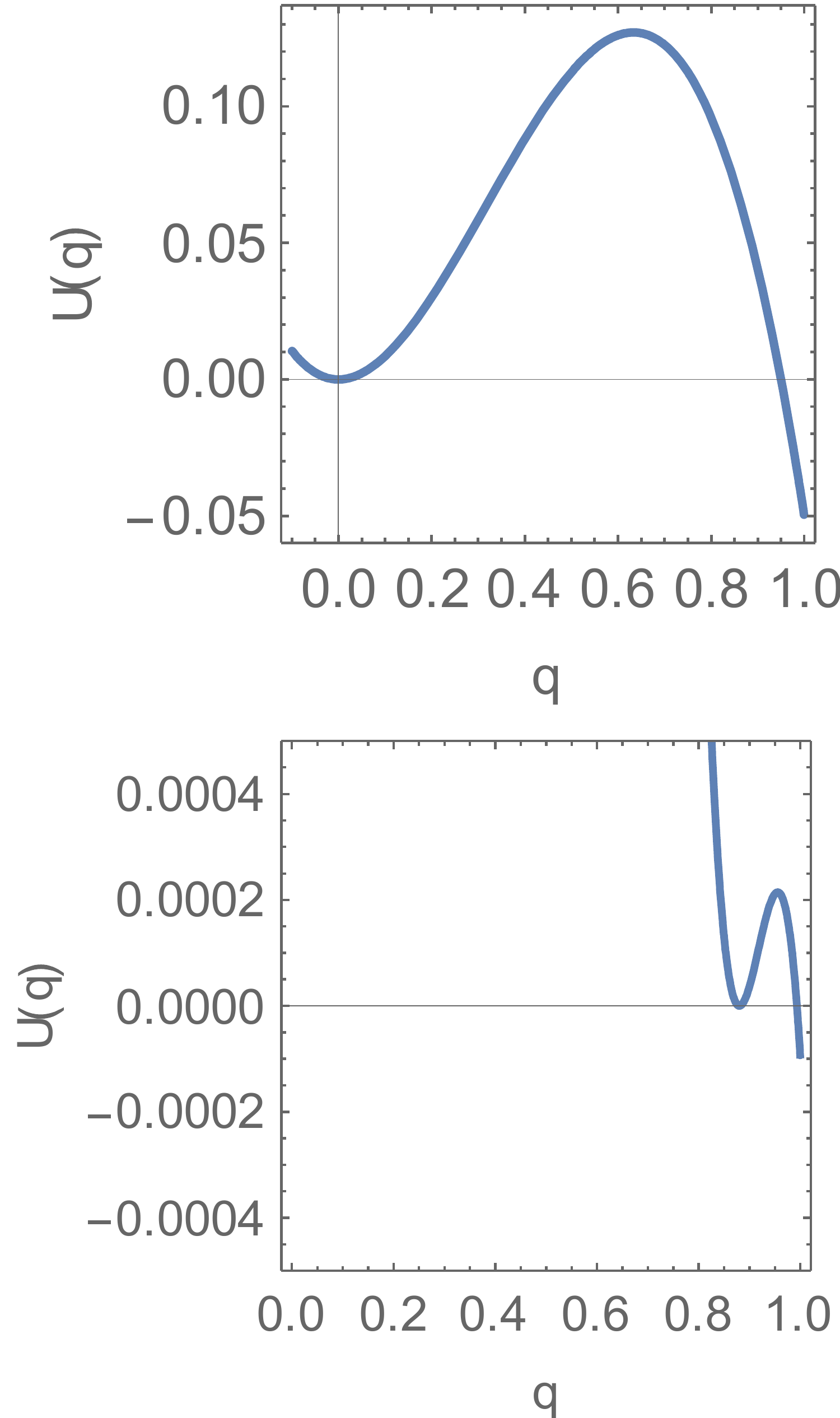} \includegraphics[width=0.67\columnwidth]{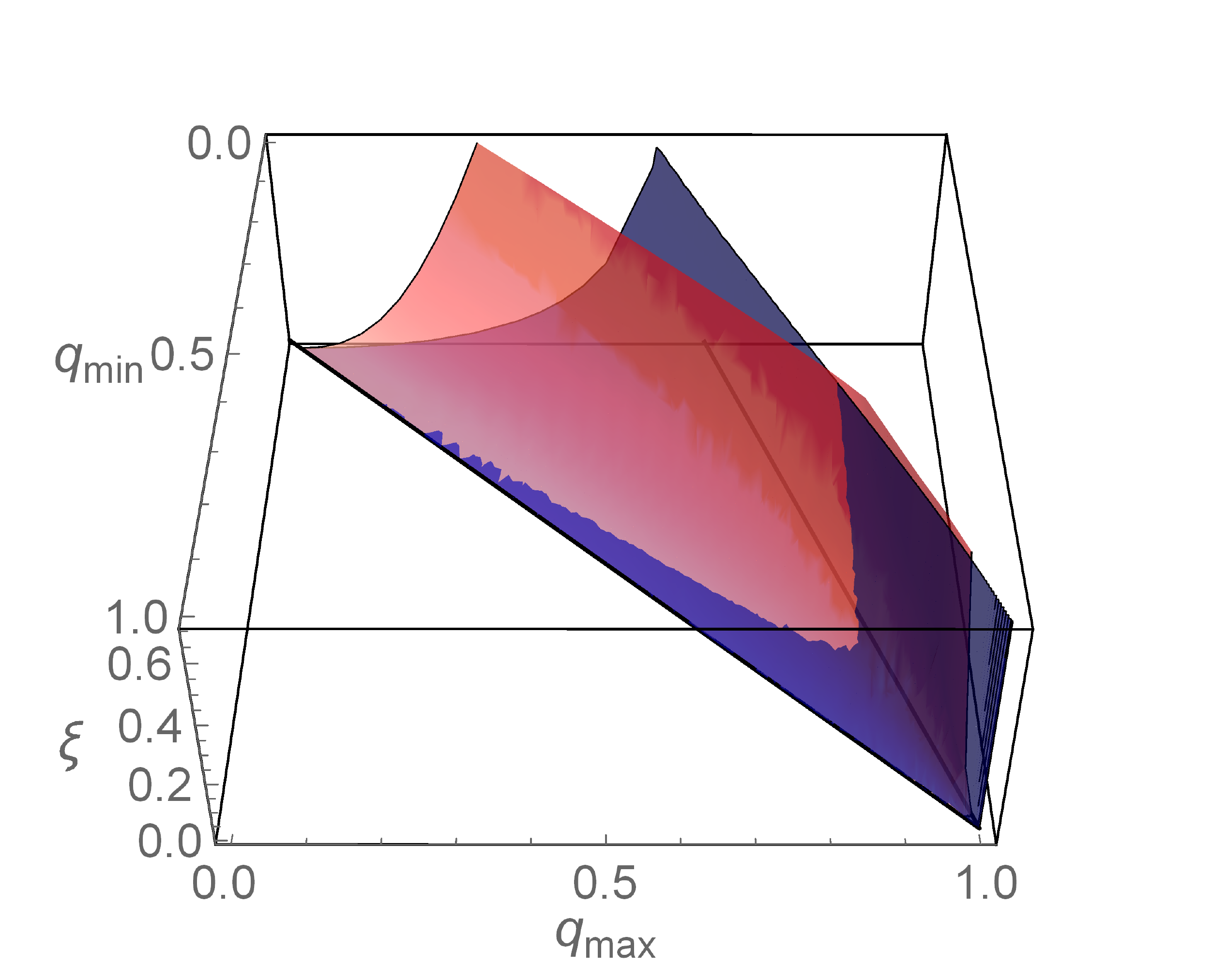}\protect\caption{\emph{Left: }Potential barrier between the metastable and the ground
state Eq.~(\ref{eq:CubicPotential}) corresponding to $s=1$ for
two pairs of values $(q_{min},q_{max})$ equal to $(0,\,0.633)$ (upper
plot), $(0.88,\,0.955)$ (lower plot). \emph{Right:} Transition action
at the optimal freezing point for QA (red) and SA (blue) as a function
the location of metastable state $q_{min}$ and the top of the barrier
$q_{max}.$ The lower of the two surfaces corresponds to the shorter
computation time. QA is advantageous for a range of parameters corresponding
to a sufficiently narrow potential barriers. Solid lines in the plane
$S=0$ correspond to $q_{min}=q_{max}$ and $q_{min}=\frac{3}{2}q_{max}$
outlining the range of possible potentials in the cubic model Eq.~(\ref{eq:CubicPotential}). }

\label{fig:3DComparison}
\end{figure}

Fig.~\ref{fig:CriticalLineAction} right demonstrates that the quantum
tunneling process may be more efficient than over-the-barrier escape.
Both classical and quantum transition rates, and therefore the corresponding
computation times, scale exponentially with the system size, $\tau\propto\exp(\alpha N)$,
yet the coefficient in the exponent is smaller in the case of QA as
compared to the classical simulated annealing, which corresponds to
a polynomial speedup. Note that it is important to compare the computation
times in Fig.~\ref{fig:CriticalLineAction} with that of the local
exhaustive in the interval $(q_{min},1)$, which is the high temperature
limit of SA. The corresponding computation time is given by the entropy
$\xi=Q_{cl}(q_{min})$, which for parameters considered in Fig.~\ref{fig:CriticalLineAction},
$Q_{cl}(0.88)\approx0.227$, is above the QA value. We further compare
the optimal performance of the open system quantum annealing $S_{opt}(\infty,s)$
to simulated annealing for a wide range of barrier shapes within the
cubic model Eq.~(\ref{eq:CubicPotential}) by varying the location
of the metastable minimum $q_{min}$ and the barrier top $q_{max}$
in Eq.~(\ref{eq:CubicPotential}), see left panel in Fig.~\ref{fig:ZeroTAction}
and Fig.~\ref{fig:3DComparison}. QA outperforms SA in a range of
the parameter space where the potential barrier separating the metastable
and the ground states is small. Note that the origin of the speedup
in the case considered here is distinct from the standard intuition
of thin and tall barriers favoring quantum tunneling, since the shape
of the potential is cubic throughout the range of parameters shown
in Fig.~\ref{fig:3DComparison}. Instead the quantum algorithm turns
out to be more efficient because it proceeds along a path with lower
entropic cost than the path that SA takes. This is a result of the
transverse field lifting the degeneracy of the metastable state in
the quantum case. Therefore the smallness of the barrier required
for the speedup in this case is determined by the comparison of the
quantum tunneling action and the SA action including entropy cost
of over-the-barrier escape. The latter being a combination of $\beta_{PT}U$
and the additional entropic cost $Q_{cl}(q_{min})-Q_{cl}(q_{max})$,
see Appendix~\ref{sec:Simulated-Annealing} for details. Note also
that the numerical value of the ratio of the logarithms of the normalized
computation times $\xi\equiv\frac{1}{N}\log\tau$ for QA and SA can
be substantial, see Fig.~\ref{fig:ZeroTAction} right.

\section{Discussion\label{sec:Discussion}}

In this paper we considered a model problem for Quantum Annealing
that allows analytical investigation and at the same time demonstrates
some key features of complex optimization problems, including the
discontinuous first order phase transition and the exponential degeneracy
of the metastable state. We demonstrate that for problems with extensive
degeneracy SA algorithm relies on over-the-barrier escape at very
low temperature $\beta\sim O(1)$ as a result the SA computation time
is exponential in $N$. At the same time we show that a computational
advantage can be gained using open system quantum annealing which
exploits the effects of thermally assisted tunneling and quantum relaxation.
The tunneling occurs between the excited states while the relaxation
within the ground state's domain of attraction brings the system down
to the lowest energy state at the end of the quantum annealing. Our
analysis demonstrates novel and counter intuitive features in the
quantum tunneling process caused by the entropy of the metastable
states, particularly, the tunneling rate decreasing with increasing
temperature. As a consequence we find that the optimal quantum annealing
regime corresponds to vanishing temperature, i.e. raising the temperature
reduces the efficiency of QA. We also find that at low temperatures
the optimal quantum tunneling rate does not always correspond to the
point of the phase transition $s_{QPT}$, in fact tunneling at $s>s_{QPT}$
can have a substantially higher rate, which can be exploited in conjunction
with noise-induced thermalization to improve the performance of the
algorithm. We find that the comparison of the quantum annealing and
simulated annealing comes down to the numerical coefficient in the
scaling of the computation time with $N$, which is determined by
the functional form of the potential barrier. We demonstrate that
optimal QA could outperform SA in a certain parameter range of our
model characterized by small potential barriers. This is in spite
of the constrained nature of the quantum tunneling process due to
the conservation laws, as opposed to the unconstrained classical Glauber
dynamics of SA. We identify an additional advantage the quantum algorithm
has over SA, which is the splitting of the degeneracy of the metastable
state by the transverse field, which means the quantum algorithm proceeds
along a trajectory with lower entropic cost as compared to SA, which
is the result of the shape of the effective potential. This suggests
a speculation that optimization problems in which entropy is a dominant
factor, yet at the same time, which are characterized by a potential
energy landscape that can be exploited for more efficient search (as
opposed to Grover's unstructured search problem~\cite{GroverOrig97})
may represent a class of problems where quantum annealing could have
a computational advantage over simulated annealing.

We also identify key features of the model affecting the performance
of the QA, specifically, the quantum fluctuations strength at the
phase transition point, and the diverging mass in the quantum kinetic
energy which both strongly affect the efficiency of quantum tunneling
in the course of the algorithm. We also find that the transverse field
term introduces an additional effective potential barrier that has
to be overcome in the course of the algorithm. In that respect it
would be interesting to use methods developed in this paper to explore
QA in mean-field models with different types of driver Hamiltonians
whose ground states are not simple product states where ferromagnetic
order competes with the transverse (XY) ferromagnetism or superfluity~\cite{Lipkin1965188,Meshkov1965199,Glick1965211,FisherFisherBoseHubbard}.

\begin{acknowledgments}
This work is supported in part by the Office of the Director of National
Intelligence (ODNI), Intelligence Advanced Research Projects Activity
(IARPA), via IAA 145483; by the AFRL Information Directorate under
grant F4HBKC4162G001. The views and conclusions contained herein are
those of the authors and should not be interpreted as necessarily
representing the official policies or endorsements, either expressed
or implied, of ODNI, IARPA, AFRL, or the U.S. Government. The U.S.
Government is authorized to reproduce and distribute reprints for
Governmental purpose notwithstanding any copyright annotation thereon.
\end{acknowledgments}



%


\appendix

\section{Simulated Annealing\label{sec:Simulated-Annealing}}

We discuss the performance of simulated annealing on the model, Eq.~(\ref{eq:Hamiltonian},~\ref{eq:CubicPotential})
in the purely classical case of $s=1$, see the rightmost panel in
Fig.~\ref{fig:U(q)}. SA algorithm is realized by starting with the
infinite temperature limit $\beta\rightarrow0$, i.e. equal occupation
of all states, and reducing the temperature to zero. For simplicity
(and without loss of generality) we consider an algorithm where \textbf{$\beta$}
changes linearly in time from $0$ to a very large value with a fixed
rate $v$.

The time it takes to find the ground state with probability approaching
$\sim1$, allowing for repeated runs of the algorithm is given, as
in the quantum case by, 

\[
\tau\sim v^{-1}\times\mathcal{P}_{GS}^{-1}.
\]
Simulated annealing relies on the system reaching thermal equilibrium
throughout at least a part of the algorithm. Such that the ground
state occupation is given by the Gibbs distribution. In problems with
a free energy barrier separating the initial and the ground state
the system's relaxation time is dominated by the over-the-barrier
escape probability with $\mathcal{F}(q)=\beta E(q)+Q_{cl}(q)$, 
\[
w(\beta)\sim\exp\left(-N\left(\mathcal{F}(q_{max})-\mathcal{F}(q_{min})\right)\right),
\]
 given by the statistical weight of the escape trajectory~\cite{HanggiRMP}
where we include the entropy of the classical state with magnetization
$q$ given by, $Q_{cl}(q)\approx-\frac{1+q}{2}\ln\frac{1+q}{2}-\frac{1-q}{2}\ln\frac{1-q}{2}.$
$w(\beta)$ reduces exponentially with decreasing temperature (growing
$\beta$) and therefore, in analogy with the quantum case considered
in the main text, there exists a freezing point $\beta=\beta_{F}$
in the course of the sweep of the inverse temperature where the relaxation
time $\sim w^{-1}(\beta_{F})$ required to achieve thermal distribution
becomes longer than the length of the algorithm,$w^{-1}(\beta_{F})\sim v^{-1}$.
After this point the values of the occupation numbers of the left
$\mathcal{P}_{L}$ and right $\mathcal{P}_{R}$ are effectively frozen.
Therefore the computation time is determined by two quantities calculated
at the freezing point, the occupation of the ground state potential
well $\mathcal{P}_{GS}=\mathcal{P}_{R}(\beta_{F}),$ and the relaxation
time at the freezing point $w^{-1}(\beta_{F})$,

\begin{equation}
\tau\approx e^{N\left(\mathcal{F}(q_{max})-\mathcal{F}(q_{min})\right)}\left(1+e^{N\left(\mathcal{F}(1)-\mathcal{F}(q_{min})\right)}\right),\label{eq:Tau}
\end{equation}
where we keep only the main order in the limit $N\rightarrow\infty$
such that $\mathcal{P}_{GS}(\beta_{F})\approx\exp(-\mathcal{F}(1))/(\exp(-\mathcal{F}(1))+\exp(-\mathcal{F}(q_{min})))$.
The optimal computation time can be obtained by minimizing with respect
to the inverse temperature at the freezing point, $\frac{\partial}{\partial\beta_{F}}\left(\frac{1}{N}\log_{2}\tau\right)=0$.
This derivative is discontinuous at the point of the phase transition
$\beta_{PT}$, where $\mathcal{F}_{R}-\mathcal{F}_{L}=0$. The computation
time is an increasing function at $\beta_{F}>\beta_{PT}$, as it is
dominated by the decreasing transition rate, the prefactor in front
of the curly brackets in Eq.~(\ref{eq:Tau}). Whereas it can be either
monotonously decreasing or increasing function at$\beta_{F}<\beta_{PT}$
depending on the competition between the prefactor and the exponents
in the brackets in Eq.~(\ref{eq:Tau}). Therefore the global minimum
of $\tau(\beta_{F})$ corresponds to the smallest value out of $\tau(\beta_{PT})$
and $\tau(0)\approx2^{N}$, in the decreasing and increasing case
respectively. The latter corresponding to the exhaustive search, i.e.
$2^{N}$ repetitions of infinitely fast SA. The ground state at $q=1$
is unique $Q(1)\approx0$ therefore the point of the classical phase
transition is at $\beta_{PT}=\frac{Q(q_{min})}{E(q_{min})-E(1)}$.
Therefore the optimal SA computation time corresponds to the smaller
of the two values,

\[
\tau\rightarrow\left[\begin{array}{cc}
\exp\left[N\left(\frac{E(q_{max})-E(q_{min})}{E(q_{min})-E(1)}Q(q_{min})+\delta Q\right)\right],\\
2^{N},
\end{array}\right.
\]
where $\delta Q\equiv Q(q_{min})-Q(q_{max})$. Note that the entropy
associated with the metastable state causes a very low transition
temperature $\beta_{PT}\sim O(1)$ and gives rise to an additional
statistical factor $\exp(\delta Q)$ slowing down the transitions
over the barrier. This additional factor appears as a prefactor in
the Kramers rate calculation~\cite{HanggiRMP}, in the model considered
here this factor is exponential and needs to be included to correctly
describe the scaling of the classical transition rate with $N$.

\section{Eigenstates overlap\label{sec:Eigenstates-overlap}}

One of the characteristics associated with complexity of a given problem
for quantum annealing algorithm is the overlap of the eigenstate wave
functions at the beginning and the end of the algorithm. We analyze
it for our model Eqs.~(\ref{eq:Hamiltonian},~\ref{eq:CubicPotential}).
The initial state $s=0$ is characterized by the maximal $x-$projection
of the total spin,

\[
\hat{S}^{x}\left|\frac{N}{2}-K\right\rangle _{x}=\left(\frac{N}{2}-K\right)\left|\frac{N}{2}-K\right\rangle _{x}.
\]
The overlap of this state with the solution of the classical problem
- state fully polarized along $z-$axis is,

\[
_{x}\left\langle N/2-K\right|\overrightarrow{0}\rangle=\left[\frac{1}{2^{N}}\left(\begin{array}{c}
N\\
K
\end{array}\right)\right]^{1/2}.
\]

\end{document}